\documentclass[journal]{IEEEtran}

\IEEEoverridecommandlockouts
\usepackage{bm}
\usepackage{cite}
\usepackage{amsmath,amssymb,amsfonts,mathrsfs}
\usepackage{algorithmic, booktabs}
\usepackage{graphicx}
\usepackage{graphics}
\usepackage[outdir=./]{epstopdf}
\usepackage{textcomp}
\usepackage[dvipsnames]{xcolor}
\usepackage{algorithm,algorithmic}[1]
\usepackage{setspace}
\usepackage{bbm}
\usepackage[bb=boondox]{mathalfa}

\usepackage{blindtext}
\usepackage{caption}
\usepackage{subcaption}
\usepackage{nomencl}
\usepackage{siunitx}
\usepackage[export]{adjustbox}
\def\BibTeX{{\rm B\kern-.05em{\sc i\kern-.025em b}\kern-.08em
T\kern-.1667em\lower.7ex\hbox{E}\kern-.125emX}}
\usepackage{etoolbox}
\usepackage{acronym}
\usepackage{multirow}
\usepackage{makecell}

\usepackage{nomencl}
\usepackage{multirow}
\usepackage{array}
\usepackage{pdflscape}
\usepackage{rotating}
\usepackage{amsthm}
\usepackage{hyperref}

\allowdisplaybreaks

\begin{document}

%\title{Wildfire Resilient Battery Energy Storage Investments under Demand Uncertainty}
\title{Wildfire Resilient Unit Commitment under Uncertain Demand %and Renewable Resource Availability 
}

% Do not provide author information in the digest. Instead provide the following information

% \author{\IEEEauthorblockN{Ryan Greenough, %Amit Harel, Mathieu Giroud, 
% Kohei Murakami, Michael Davidson, Jan Kleissl, Adil Khurram }
% \IEEEauthorblockA{Department of Mechanical and Aerospace Engineering,\\ 
% University of California, San Diego, CA 92037\\
% Email: {\{rgreenou, %aharel, mgiroud, 
% k1murakami, mrdavidson, jkleissl, akhurram\}}@ucsd.edu}}

\author{Ryan Greenough,
Kohei Murakami, Michael R. Davidson, Jan Kleissl, and Adil Khurram 
	\thanks{Ryan Greenough, Kohei Murakami, and \hbox{Adil Khurram} are with the Department
of Mechanical and Aerospace Engineering, University of California at San
Diego, La Jolla, CA 92093 USA (e-mail: rgreenou@ucsd.edu; kmurakami@epri.com; akhurram@ucsd.edu)} 
\thanks{
Michael R. Davidson is with the  Department of Mechanical and Aerospace Engineering and the School of Global Policy and Strategy, University of California at San Diego, La Jolla, CA 92093 USA (email: mrdavidson@ucsd.edu)}
\thanks{
Jan Kleissl is with the Center for Energy Research, Department of
Mechanical and Aerospace Engineering, University of California at San
Diego, La Jolla, CA 92093 USA (e-mail: jkleissl@ucsd.edu).}}

\maketitle

\makenomenclature

\begin{abstract}
Public safety power shutoffs (PSPS) are a common pre-emptive measure to reduce wildfire risk due to power system equipment failure.
% mitigate a power system's operation from contributing to wildfire spread. 
System operators use PSPS to de-energize electric grid elements that are either prone to failure or located in regions at a high risk of experiencing a wildfire. Successful power system operation during PSPS involves coordination across different time scales. Adjustments to generator commitments and transmission line de-energizations occur at day-ahead intervals while adjustments to load servicing occur at hourly intervals. Generator commitments and operational decisions have to be made under uncertainty in electric grid demand and wildfire potential forecasts. 
% Power system optimization methods must consider this uncertainty of wildfire and demand parameters and the time dependence of real-time operation on day-ahead unit commitments and line de-energizations. 
% Given the high impact and low frequency of wildfire disasters, operators must consider a range of operating scenarios that could be costlier than typical operations. 
%This paper presents a framework for system operators to optimize generator commitment and line de-energizations during PSPS planning and operation. 
This paper presents deterministic and two-stage mean-CVaR stochastic frameworks to show how the likelihood of large wildfires near transmission lines affects generator commitment and transmission line de-energization strategies. The optimal costs of commitment, operation, and lost load on the IEEE 14-bus and 24-bus test systems are compared to the costs generated from prior optimal power shut-off (OPS) formulations. The proposed mean-CVaR stochastic program generates less total expected costs evaluated with respect to higher demand scenarios than costs generated by risk-neutral and deterministic methods. 
%The proposed mean-CVaR stochastic program empirically shows cost savings of optimizing the load served in anticipation of high system demands.
\end{abstract}


\mbox{}

\nomenclature[]{\(\mathcal{B}\)}{Set of buses (nodes)}
\nomenclature[]{\(B\)}{Number of buses}

\nomenclature[]{\(\mathcal{L}\)}{Set of transmission lines (edges)}

\nomenclature[]{\(\mathcal{U}\)}{Undirected graph}
\nomenclature[]{\(H\)}{Number of samples in the DA optimization horizon}
\nomenclature[]{\(B_{ij}\)}{System susceptance value connecting bus $i$ to bus $j$}
\nomenclature[]{\(\theta_{i,t,\omega}\)}{Phasor angle at bus $i \in \mathcal{N}$ and scenario $\omega$}
\nomenclature[]{\(\underline{p}_{g},\overline{p}_{g}\)}{Lower/upper active generation limit for each generator $g \in \mathcal{G}$}
\nomenclature[]{\(\underline{U}_{g},\overline{U}_{g}\)}{Lower/upper active generation ramp limit for each node $i \in \mathcal{G}$}

\nomenclature[]{\(\Delta t\)}{Incremental time step in the optimization horizon}

\nomenclature[]{\(p_{g,t,\omega}\)}{Active generation provided by each generator $g \in \mathcal{G}$ at time $t$ and scenario $\omega$}

\nomenclature[]{\(D_{\text{Tot},\omega}\)}{Total load served for each scenario $\omega$}
\nomenclature[]{\(p_{d,t,\omega}\)}{Nodal demand (uncertain) at each node $d \in \mathcal{D}$ at time $t$ and scenario $\omega$}

\nomenclature[]{\(C^{\mathrm{VoLL}}_{d}\)}{Value of Lost Load for each demand node $d \in \mathcal{D}$}
\nomenclature[]{\(x_{d,t,\omega}\)}{Fraction of demand served for $d \in \mathcal{D}$, at time $t$ and scenario $\omega$}
\nomenclature[]{\(\mathcal{H}\)}{Set of time-steps in the optimization horizon}
\nomenclature[]{\(\mathcal{G}\)}{Set of generators}
\nomenclature[]{\(\mathcal{D}\)}{Set of demands}

\nomenclature[]{\(R_{ij,t}\)}{Wildfire risk of a transmission line measured}
\nomenclature[]{\(R_{\text{Tot}}\)}{Cumulative system transmission line wildfire risk}
\nomenclature[]{\(R_{\text{slack}}\)}{Wildfire risk slack variable}
\nomenclature[]{\(\overline{R}_{\text{slack}}\)}{Upper bound on violation of wildfire risk tolerance equality constraint}
\nomenclature[]{\(z_{g,t},z_{ij,t}\)}{Active status for generators and transmission lines}

\nomenclature[]{\(p^{\text{aux}}_{g,t,\omega}\)}{Auxiliary variable for a generator $g \in \mathcal{G}$ at time $t$ and scenario $\omega$ whose definition prevents violation ramping constraints when generator $g$ is turned off}
\nomenclature[]{\(z^{\text{up}}_{g,t},z^{\text{dn}}_{g,t}\)}{Startup and shutdown binary decisions}
\nomenclature[]{\(t^{\text{MinUP}}_{g},t^{\text{MinDn}}_{g}\)}{Minimum up and down times for each generator $g \in \mathcal{G}$}
\nomenclature[]{\(\underline{\theta},\overline{\theta}\)}{Lower/upper phase angle difference}
\nomenclature[]{\(p_{ij,t,\omega}\)}{Active power flow through transmission line from bus $i$ to bus $j$ at time $t$ and scenario $\omega$}
\nomenclature[]{\(\underline{p}_{ij},\overline{p}_{ij}\)}{Lower/upper thermal limits on transmission lines on flow $i$ to $j$ at time $t$}

\nomenclature[]{\(\beta\)}{Risk-averseness trade-off parameter}
\nomenclature[]{\(\omega\ \)}{A demand scenario in a given set of scenarios $\Omega$}
\nomenclature[]{\(\Omega,\Omega^{\text{RT}}\)}{Set of day-ahead and real-time demand scenarios}

\nomenclature[]{\(\Pi_{\omega}\)}{Economic costs for a given demand scenario $\omega$}
\nomenclature[]{\(\Pi^{NoUC}_{\text{det}}\)}{Operating costs for a deterministic SPSPS/OPS}
\nomenclature[]{\(\overline{\Pi}^{NoUC}_{\text{det}}\)}{Maximum operating costs in OPS objective}
\nomenclature[]{\(\overline{R}_{\text{Tot}}\)}{Maximum cumulative wildfire risk in OPS objective}
\nomenclature[]{\(\alpha\)}{Weight balancing between economic costs and total wildfire risk}
\nomenclature[]{\(\pi_{\omega}\)}{Probability of each demand scenario $\omega$}
\nomenclature[]{\( \mathbb{E}[ \cdot]\)}{Expected Value}
\nomenclature[]{\(\nu\)}{$\text{CVaR}_{\epsilon}$ auxiliary variable representing the Value at Risk ($\text{VaR}_{\epsilon}$) at optimum}
\nomenclature[]{\(\gamma_{\omega}\)}{Auxiliary variable for each demand scenario $\omega$ representing the shortfall between $\nu$ and $\Pi_{\omega}$}
\nomenclature[]{\( \text{CVaR}_{\epsilon}\)}{Conditional Value at Risk at confidence level $\epsilon$}
\nomenclature[]{\(C_g\)}{Variable cost of generation for each $g \in \mathcal{G}$}
\nomenclature[]{\(C_{g}^{\mathrm{dn}},C_{g}^{\mathrm{up}}\)}{Cost to shutdown/startup a generator $g \in \mathcal{G}$}
\nomenclature[]{\(\|\cdot\|\)}{Cardinality of Set}
\nomenclature[]{\(f^{\mathrm{uc}},f^{\mathrm{oc}},f^{\mathrm{VoLL}}\)}{Unit commitment (uc), operating (oc), and VoLL costs}
\nomenclature[]{\(R_{\mathrm{tol}}\)}{An operator's WFPI (unitless) risk tolerance level }
\nomenclature[]{\(\Phi_{\omega}\)}{Operating costs for a given demand scenario $\omega$ }
\nomenclature[]{\(\xi\)}{Realization of random data}
\nomenclature[]{\(\Xi\)}{Set of random data realizations}
\nomenclature[]{\(\epsilon\)}{Confidence level for $\text{CVaR}_{\epsilon}$}

\printnomenclature

\section{Introduction}
\label{section:Introduction}
Since 2012, an important short-term strategy for electric utilities to proactively reduce wildfire ignition probabilities is public safety power shut-offs (PSPS) \cite{Muhs,Arab,CPUC}. A PSPS seeks to ensure reliable electrical grid operation during wildfires~\cite{Arab} and maximize load served while minimizing wildfire risk. In \cite{Abatzoglou}, a significant increase in PSPS is predicted due to drier autumn seasons in California's future. Across the US, the annual wildfire frequency is predicted to increase by 14\% by 2030 and 30\% by 2050 ~\cite{EPA, UN}. 

In practice, system operators typically rely on experience and rule-based heuristics to schedule PSPS, such as the transmission line and area wildfire risk heuristics \cite{Rhodes}. 
% The area heuristic de-energizes all components within a predefined area that exceeds a maximum area risk threshold. The transmission line heuristic de-energizes individual lines that exceed a maximum line risk threshold and other system components that have been isolated from generation sources after line shut-offs \cite{Rhodes}. A notable drawback to the transmission line and area heuristics is that high wildfire-risk lines that are critical to helping serve a large portion of the demand may be shut down. 
\cite{Rhodes} shows that because both of these wildfire risk-based heuristics do not consider how network de-energizations affect optimal power flows in the network, more system wildfire risk is present for the same demand served on networks de-energized by field heuristics than by using their optimal power shut-off (OPS) framework. 

% Their OPS model addresses the drawbacks of not considering the relationship of line-shut-offs and load delivery by optimizing for a weighted trade-off between wildfire risk and load delivery when subjected to optimal power flow constraints. 

% Their OPS model decreases the amount of system wildfire risk for the same amount of load served as the field heuristic. 

% \cite{Rhodes} demonstrate this improvement in pareto optimal load served wildfire risk trade-off points in simulations on the Reliability Test System Grid Modernization Lab Consortium (RTS-GMLC) transmission grid using data loosely based on 2019 test data from Southern California at a snapshot in time. 

The OPS framework has been extended in several directions to include additional planning and investment decisions that would affect PSPS strategies~\cite{Kody, Rhodes2, Kody2, Umunnakwe, Bayani}. 
\cite{Kody} optimizes investments in distributed energy resource (DER) infrastructure (e.g. distributed solar and wind) and line hardening measures to reduce cumulative system wildfire risk and improve load delivery in regions of the RTS-GMLC and WECC more vulnerable to PSPS events. \cite{Rhodes2} incorporates $N$-$1$ security contingencies to robustify PSPS operation to unexpected line failures. \cite{Kody2} considers how de-energizations can disproportionately affect certain areas of the grid and suggests that operators should add fairness functions, such as the minimizing maximum load shed and minimizing a weighted penalized load shed, to PSPS planning to spread the burden of load shed and de-energizations across the grid. 
% \cite{Kody2} recommends a thorough analysis of the load shedding and wildfire risk mitigation outcomes before implementation because certain fairness metrics can lead to unnecessary increases in local load shedding.

% \cite{Astudillo, Umunnakwe, Bayani} make contributions related to the forecasting of wildfire risk metrics. \cite{Astudillo} constructs a time-varying wildfire risk by temporally scaling the static wildfire risk values used in \cite{Rhodes} with changes in temperature and nodal demands. \cite{Umunnakwe} develops a wildfire risk model to predict ignition probabilities and fire spread via a neural network.
% \cite{Bayani} contributes a wildfire forecasting surrogate that approximates the non-linear droop dynamics of transmission lines.  

% Despite adjusting OPS objectives and constraints to account for unplanned outages, renewable infrastructure, and fair load shedding, in ~\cite{Rhodes, Kody, Rhodes2, Kody2, Umunnakwe} OPS may choose more costly commitment and operational strategies because OPS does not consider the unit commitment and operation costs.
However, the OPS defined in ~\cite{Rhodes, Kody, Rhodes2, Kody2, Umunnakwe} does not consider commitment and operational costs and may choose more costly commitment and operational strategies. In OPS, developing appropriate weights for trading off normalized load loss versus normalized wildfire risk relies heavily on operator experience. Instead, we propose framing all objectives in terms of economic costs. Furthermore,  the OPS weighted objective defined in ~\cite{Rhodes, Kody, Rhodes2, Kody2, Umunnakwe} does not allow the operator to find all Pareto optimal points for demand and wildfire risk because the total wildfire risk objective is non-convex.

% and could lead to difficulty when interpreting results.
% incorporating complex engineering and social constraints or wildfire forecasting models to analyze the effects of. By assuming that operators seek pareto optimal trade-offs between the cumulative wildfire risk of energized grid-connected components (loads, transmission lines, generators, stationary storage units, etc.)  and the total load served. And \cite{Umunnakwe} only considers the value of lost load in the objective function. Strategies from \cite{Rhodes, Kody, Rhodes2, Kody2, Astudillo, Umunnakwe}   
Similar to the present work, \cite{Bayani} considers generator production costs in the deterministic objective function and monetizes line shut-offs; however, \cite{Bayani} does not consider any time-coupled constraints to model the scheduling limitations of generator dispatch and line de-energizations. \cite{Bayani} does not include ramping and commitment constraints, delays in restarting transmission lines after damages, and generator commitment costs. 

Furthermore, the optimization models in~\cite{Rhodes, Kody, Rhodes2, Kody2, Umunnakwe, Bayani} solve only the deterministic version of the PSPS by assuming perfect forecasts of both nodal demands and wildfire risk of grid components. \cite{Moreno} proposes a probabilistic DER planning model for PSPS events in Chile by minimizing the expected costs of investment, operation, and unserved energy. 
% The optimization includes preventive measures (including investments in DER infrastructure and contracts for an appropriate volume of demand) and corrective measures (including curtailment of demand and dispatch of mobile DERs). 
 They note that there is large uncertainty in the sampled probability distribution of the outage scenarios used in the stochastic optimization. This uncertainty could drive away risk-averse investors because they may be more interested in viewing scenarios that generate worse-case profits rather than minimizing expected costs from a much larger selection of possible outage scenarios. \cite{Moreno} left studying the impacts of investor risk aversion to high planning and operating costs due to uncertain outage scenarios for future work. %\cite{Trakas1,Trakas2} 
\cite{Trakas2} constructs a robust unit commitment problem to model resilient operations during extreme weather events. 
%\cite{Trakas1} only explores minimizing against the worst-case realizations of transmission line damages due to high winds. And there is no consideration for uncertainty in any parameters that affect operational decisions, such as demand or resource availability. 
\cite{Trakas2} considers uncertainties in operational parameters and explores how wildfire spread can affect the thermal limits of transmission lines. However, \cite{Trakas2} takes a risk-neutral approach to minimizing operational costs and does not consider how a risk-averse approach could alter line de-energizations and unit commitment decisions. 

In this work, a mean-risk two-stage stochastic version of the PSPS (SPSPS) is formulated. A risk-averseness parameter $\beta$ trades-off between minimizing the total expected costs with the $\text{CVaR}_{\epsilon}$ costs. Uncertainty is introduced in the maximal demand served for each bus in the network. The risk averseness parameter incorporates an operator's tendency to reduce spending when demand scenarios generate higher total economic costs (i.e. sum of the commitment and operational costs). Wildfire risk is represented by the Wildland Fire Potential Index (WFPI) and daily WFPI forecasts from the USGS are mapped to each transmission line. Our framework discourages operating transmission lines in fire-prone regions via a tolerance level set by the operator, $R_{\textrm{tol}}$. %WFPI-based Large Fire Probability (WLFP) 
%via regression decision trees. 
Real-time optimized device shut-offs and power injections are determined on a rolling horizon basis for the IEEE 14 bus %and RTS-GMLC 
transmission grid.

Our main contribution is modeling the non-anticipatory nature of preemptive commitment and line de-energizations performed during PSPS events and provide operators with the ability to be risk-averse to high economic costs. All prior work assumes operators are risk-neutral toward different cost scenarios. Several works assume operators can perfectly anticipate demands (\cite{Rhodes, Kody, Rhodes2, Kody2, Bayani}) %,Trakas1) 
or instantaneously adjust generator commitment status (\cite{Rhodes, Kody, Rhodes2, Kody2,  Umunnakwe, Bayani,Trakas2}).  To overcome this gap in the literature, we propose a two-stage stochastic day-ahead unit commitment model that incorporates uncertainties in demand forecasts and different timescales of operation while minimizing shut-down and start-up generation costs, operational costs, and the cost of unserved demand subject to a predefined level of risk for transmission line damages due to nearby wildfires. Finally, because we define wildfire risk as a constraint rather than an objective, our optimization can find all Pareto optimal points given a careful selection of the $R_{\text{tol}}$ parameter \cite{Haimes}.
% The non-anticipatory and minimum up and down time constraints in generator commitments and line de-energization decisions more accurately model the preemption of PSPS strategies than ~\cite{Rhodes, Kody, Rhodes2, Kody2, Astudillo, Umunnakwe, Bayani,Trakas2} without those constraints. 

%\item Utilize a forecast of WFPI values for each transmission line generated from a decision tree regression rather than downloaded directly forecasts from the US Geological Survey (USGS) database for less conservative wildfire risk minimization

% \item Illustrate savings in expected costs from generator commitment and line de-energization decisions generated from a two-stage stochastic framework as compared to costs from commitment and de-energization decisions generated based on the expected demand when the network has been reduced in size due to line de-energizations. This cost savings justifies the usage of a two-stage stochastic unit commitment framework as opposed to a deterministic unit commitment.  

Applying the model we are able to generate several novel findings and improved outcomes:
\begin{enumerate}
\item As opposed to deterministic strategies that are both popular in the literature, such as the OPS in \cite{Rhodes, Kody, Rhodes2, Kody2} and widely used in the field, we include commitment and operational costs into the optimization, which reduces total expected costs.

%that also incorporates uncertainties in the demand via demand scenarios %, wind \& solar availability through a two-stage stochastic optimization framework when compared to a deterministic optimization based on the expected value of demand.

\item We demonstrate that risk-averse operators can reduce the costs of operation in high-demand scenarios but at the possible expense of losing more demand at buses to blackouts. 

% Prior work did not give operators the flexibility to choose a de-energization and commitment strategy that may be more costly in expected value but less costly in the event of higher demand scenarios than a risk-neutral approach. 
%risk averseness to demand uncertainty changes transmission line shut-down and load shedding decisions

% Commitment and operational costs are critical components in the total costs of PSPS decisions; ~\cite{Rhodes, Kody, Rhodes2, Kody2, Astudillo, Umunnakwe} focused only on serving customer demand and mitigating wildfire risk operationally.
\end{enumerate}

The rest of the paper is organized as follows. Section~\ref{section:Modeling of Public Safety Power Shut-offs for Wildfire Risk Mitigation} introduces the PSPS formulations. Section~\ref{section:Results} details results for the deterministic and stochastic PSPS unit commitment in wildfire-prone regions on the IEEE 14-bus system. %Adjustments in battery capacity and location empirically show the limits of Pareto Optimal tradeoffs between wildfire risk mitigation and total load served. In Section \ref{Tentative Modeling of the Effect of EV Penetration on Grid Resilience}, we will include mobile storage units when determining the optimal portfolio of wildfire-resilient DERs for PSPS, develop a two-stage stochastic PSPS optimization that captures an operator's risk-averseness to regional demand uncertainties, and investigate decentralized control strategies that help DER owners minimize costs of commitment and operation given a maximal allowable amount risk from transmission line energization.
%These approaches are demonstrated on an IEEE 14-bus system. %and the RTS GMLC. 
Section~\ref{section:Conclusion} summarizes and concludes the paper.

\section{Modeling Public Safety Power Shut-offs for Wildfire Risk Mitigation}
\label{section:Modeling of Public Safety Power Shut-offs for Wildfire Risk Mitigation}
\begin{figure}[ht]
\centerline{\includegraphics[width=\columnwidth]{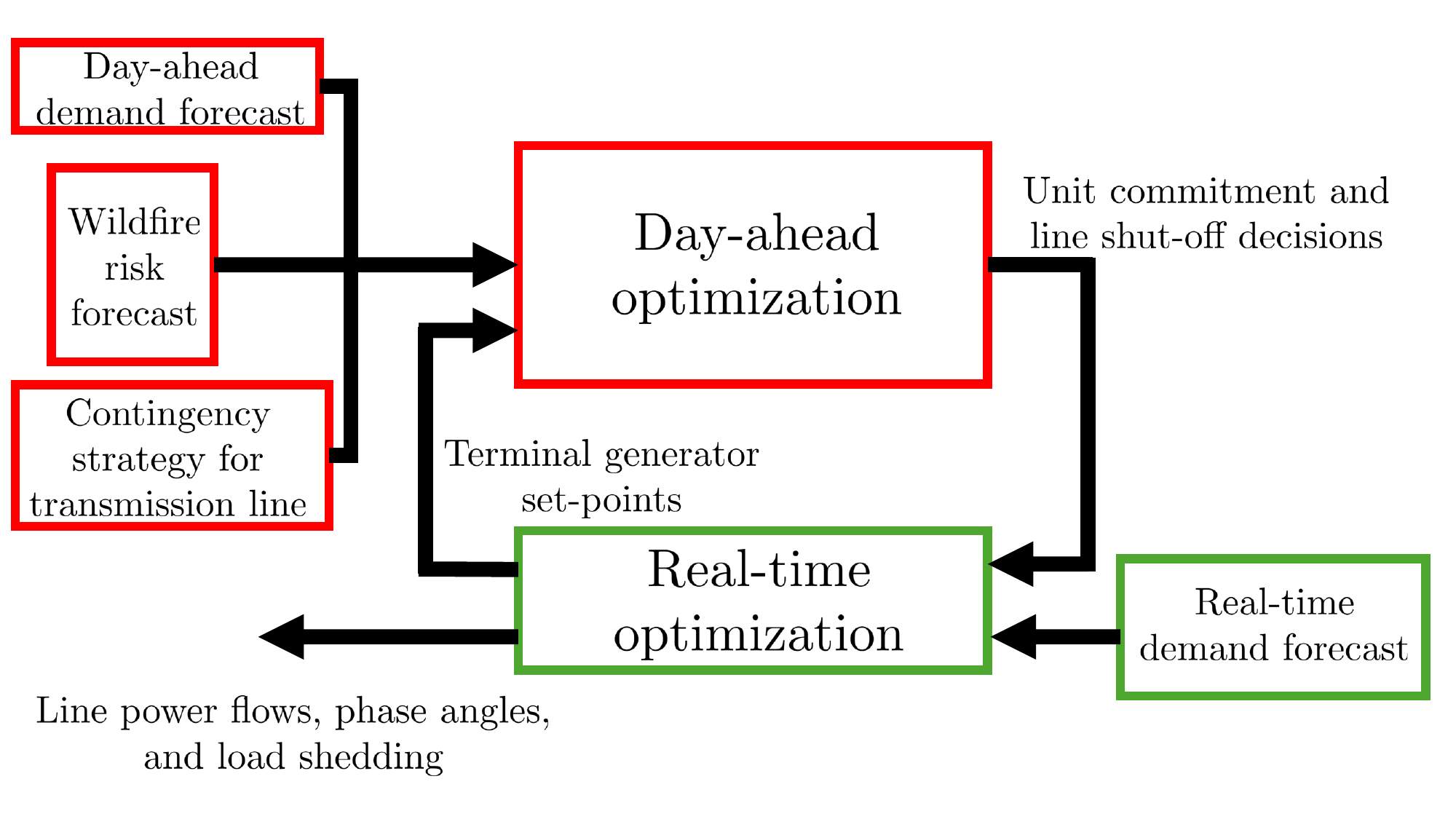}}
\caption{Block diagram showing the data inputs and decision outputs for each stage of the two-stage stochastic optimization. The red border highlights the inputs for the Day-ahead optimization while the green border highlights the inputs for the real-time post-optimization test.}
\label{fig:OptBlockDiagram}
% OneDrive File: C:\Users\rgree\OneDrive - UC San Diego\Block Diagrams for Wildfire Journal Paper.pptx
\vspace{-2em}
\end{figure}

% This section presents the objective function and constraints used to formulate the proposed two-stage stochastic PSPS problem. 
A two-stage problem is presented in which operators first make day-ahead decisions, such as generator commitments and transmission line de-energizations, then make real-time adjustments to supply and demand mismatches through operational decisions, including load shedding and generation adjustments. Day-ahead decisions are influenced by day-ahead demand and wildfire risk forecasts and real-time decisions are influenced by realizations of real-time demand (see Figure~\ref{fig:OptBlockDiagram}). 
% Inputs and outputs of each stage of the optimization are summarized in Figure \ref{fig:OptBlockDiagram}.  

% In Sections \ref{subsection:Stochastic Objective Function} through \ref{subsection:Two Stage Stochastic Problem}, all components of the two-stage stochastic PSPS are detailed including the objective function (Section \ref{subsection:Stochastic Objective Function}), unit commitment constraints (Section \ref{subsection:Unit Commitment Constraints}), transmission line contingency constraints (Section \ref{subsection: Constraint for Enacting No Less Than k Damaged Lines}), and operational constraints (Section \ref{subsection: Operational Constraints}). In Section \ref{subsection: Deterministic Formulation}, we present the deterministic PSPS problem as a special case of the stochastic program by using the mean prediction of fire risk values determined from the wildfire prediction model and the mean across scenarios of the demand forecasts. This model optimizes commitment costs, operational costs, and slack in the allowable amount of wildfire risk determined by the system operator. Then we describe common benchmarks used to determine the effectiveness of our PSPS strategy against other PSPS strategies presented in ~\cite{Kody, Rhodes2, Kody2, Astudillo, Umunnakwe, Bayani}. In Section \ref{subsection:Two Stage Stochastic Problem}, we introduce the formulations for the first and second stages and motivate the need for a stochastic formulation of the PSPS. % and explain findings resulting from the uncertainty in the demand. % and availability of renewable generation.
\subsection{Preliminaries}
Let $\mathcal{P}=(\mathcal{B}, \mathcal{L})$ be the graph describing the power grid where $\mathcal{B}=\{1, \dots, B\}$ is the set of $B$ buses in the network, and $\mathcal{L}$ is the set of edges such that two buses $i,j \in \mathcal{B}$ are connected by a transmission line if $(i,j)\in \mathcal{L}$. The set of buses with generators and loads are collected in $\mathcal{G}$ and $\mathcal{D}$ respectively, and $\mathcal{H}=\{1, \dots, H\}$ is the set of time indices over the horizon $H$ of the optimization problem. A DC-OPF is used to approximate the line power flows and bus power injections; all references to power are to its active power. At any timestep $t$, the power injected by the generator at bus $g \in \mathcal{G}$ in scenario $\xi \in \Xi$ is denoted by $p_{g,t,\xi}$. Similarly, $p_{d,t,\xi}$ is the load at bus $d \in \mathcal{D}$. 
% Boldface variables represent vector quantities; $\bm{p_{g}}\in\mathbb{R}^{|\mathcal{G}||\mathcal{H}|}$ is a collection of all power injected by for every generator  $g \in \mathcal{G}$ at every time timestep $t \in \mathcal{H}$.
Power flowing through the line $(i,j) \in \mathcal{L}$ in scenario $\xi$ is $p_{ij,t,\xi}$ and the phase angle at bus $i\in \mathcal{B}$ is denoted by $\theta_{i,t,\xi}$. Finally, the binary variables are denoted by $z \in \{0,1\}$ with an appropriate subscript to capture component shut-off decisions. % in the next subsection. 
%Any function with $xi$ as an input, such as, $f_{\text{VoLL}}(\bm{x_d}, \bm{p_d}, \bm{\xi})$ informs the read that the function's output is a random scalar. While any function with the scenario of the random data $\xi_{\omega}$ as an input, such as $\Pi_{\omega}(\cdot,\xi_{\omega})$, informs the reader that the function's output is a deterministic scalar.
\subsection{Objective Function}
\label{subsection:Stochastic Objective Function}
The objective of the stochastic PSPS with unit commitment is to minimize the cost of starting and shutting off generators ($f^\text{uc}$), operating costs ($f_{\xi}^\text{oc}$) of generators, and the cost associated with the fraction of load shed ($f_{\xi}^{\text{VoLL}}$), called the value of lost load (VoLL) defined as % in~\eqref{eq:STOUC}-\eqref{eq:STOVoLL} below, 
\begin{align}
{f^{\text{uc}}}(\bm{z^{\text{up}}_{g}},\bm{z^{\text{dn}}_{g}}) &= \mathop \sum \limits_{t \in \mathcal{H}} \left (\mathop \sum \limits_{g \in \mathcal{G}} {C_g^{\text{up}}{z^{\text{up}}_{g,t}} + C_g^{\text{dn}}{z^{\text{dn}}_{g,t}}} \right ), \label{eq:STOUC}\\ 
 f_{\xi}^{\text{oc}}(\bm{p}_{\bm{g},\xi}) &= \mathop \sum \limits_{t \in \mathcal{H}} \left(\sum \limits_{g \in \mathcal{G}} {C_g p_{g,t,\xi}} \right), \label{eq:STOOC} \\
{f_{\xi}^{\text{VoLL}}}(\bm{x_d}, \bm{p}_{\bm{d},\xi}) & = \mathop \sum \limits_{t \in \mathcal{H}} \left(\mathop \sum \limits_{d \in \mathcal{{D}}} C_{d}^{\text{VoLL}}\left(1-x_{d,t,\xi}\right)p_{d,t,\xi} \right), \label{eq:STOVoLL}
\end{align}
where $C_g^{\text{up}}$ and $C_g^{\text{dn}}$ are the generator start up and shut down cost which together with the binary variables $z_{g,t}^{\text{up}}$ and $z_{g,t}^{\text{dn}}$ capture the one-time cost incurred when bringing a generator online or offline. $C_g$ is each generator's associated marginal cost. Only demand uncertainty is considered which is captured in $\xi$ and the corresponding demand is given by $x_{d,t,\xi} \in [0,1]$ and $C_d^{\text{VoLL}}$ is the cost incurred as a result of shedding $(1-x_{d,t,\xi})$ proportion of the load, $p_{d,t,\xi}$. Let $p_{g,\xi} = (p_{g,1,\xi}, \dots, p_{g,H,\xi})^\top$ be the vector of $p_{g,t,\xi}$ for generator $g$ given uncertainty $\xi$, then $\bm{p}_{\bm{g},\xi}$ for all generators are denoted as $\bm{p}_{\bm{g},\xi} = (p_{1,\xi}, \dots, p_{G,\xi})^\top$. The variables $\bm{p}_{\bm{d},\xi}$, $\bm{x}_{\bm{d},\xi}$, $\bm{z_g}^{\text{up}}$,  $\bm{z_g}^{\text{dn}}$ are defined in a similar manner. The resulting objective function is given by,
\begin{align} 
\Pi_{\xi} = f^{\text{uc}}(\bm{z^{\text{up}}_{g}},\bm{z^{\text{dn}}_{g}})+f_{\xi}^{\text{oc}}(\bm{p}_{\bm{g},\xi}) + f_{\xi}^{\text{VoLL}}(\bm{x}_{\bm{d},\xi},\bm{p}_{\bm{d},\xi})+ R_{\text{slack}},\label{eq:STOObj}
\end{align}
where $R_{\text{slack}}$ is a design parameter, defined in Section~III-E, and is added to prioritize low-risk line shut-off strategies. Due to the uncertainty $\xi$, the objective function is stochastic, therefore, the risk-neutral PSPS aims to minimize the expected value of $\Pi_{\xi}$. A scenario-based two-stage approach is used to solve the resulting stochastic optimization problem as described in Section~III.F. %, the two-stage method is developed to solve the resulting stochastic optimization problem.
\vspace{-0.8em}
\subsection{Unit Commitment Constraints}
\label{subsection:Unit Commitment Constraints}
The unit commitment constraints \eqref{eq:UC1}-\eqref{eq:UC2} are used to enforce minimum up time ($t^{\text{MinUp}}_{g}$) and down time ($t^{\text{MinDn}}_{g}$) of generators. Similarly, the constraint~\eqref{eq:UC3} guarantees consistency between the binary variables $z^{\text{up}}_{g,t}$ and $z^{\text{dn}}_{g,t}$.
% are startup/shutdown binary decisions of a generator $g$  at time $t^{\prime}$,  is the minimum time after start-up before a unit can shut down and $t^{\text{MinDown}}_{g}$ is the minimum time after shut-down before a unit can start again. 
In all simulations, it is assumed that all generators are initially off.
\begin{subequations}
\begin{align}
z_{g,t} &\geq \sum_{t^{\prime} \geq t-t^{\text{MinUp}}_{g}}^{t} z^{\text{up}}_{g, t^{\prime}}, &\forall g \in \mathcal{G}, \: t\in \mathcal{H} \label{eq:UC1} \\
1-z_{g,t} &\geq\sum_{t^{\prime} \geq t-t^{\text{MinDn}}_{g}} z^{\text{dn}}_{g, t^{\prime}},  &\forall g \in \mathcal{G}, \: t\in\mathcal{H} \label{eq:UC2}\\
z_{g, t+1}-z_{g, t}&= z^{\text{up}}_{g, t+1}-z^{\text{dn}}_{g, t+1}, &\forall g \in \mathcal{G}, \: t\in \mathcal{H} \label{eq:UC3}
\end{align}
\end{subequations}

\subsection{Operational Constraints}
\label{subsection: Operational Constraints}

The power of generator $g$ is limited as 
\begin{equation}
    z_{g,t} \underline{p}_{g} \le p_{g,t,\xi} \le z_{g,t} \overline{p}_{g}, \: \forall t \in \mathcal{H}, \: g \in \mathcal{G},\: \forall \xi \in \Xi,
    \label{eq:Pg}
\end{equation}
where $\underline{p}_{g}$ and $\overline{p}_{g}$ are the minimum and maximum power generation limits of generator $g$. 
% Ramp rate constraints are also included in the optimization problem, given by, 
% This constraint ensures that generation at a generator, $g$, goes to 0 if the bus at the same node is de-energized and stays within physical limits during normal operation.  
% Constraint \eqref{eq:Pg} restricts each generator by the upper and lower limits on their outputs at each hourly period, which is consistent with the generator capacity constraints in \cite{Rhodes} and \cite{Kody}. 
% We also add more realistic constraints including ramp rates and minimum up and down time constraints to show more realistic trade-offs between wildfire risk to total load served. The typical ramping constraint, detailed in Constraint \eqref{eq:StandardRamp}, states that the difference between two consecutive generator outputs $p_{g, t+1,\omega}$ and $p_{g, t,\omega}$ must be bounded below and above by minimum and maximum ramping values $U^{\min}_{g,t}$ and $U^{\max}_{g,t}$,
% \begin{align}
% \label{eq:StandardRamp}
% \underline{U}_{g} &\leq p_{g,t+1}-p_{g,t} \leq \overline{U}_{g} \quad \forall g \in \mathcal{G}, t\in \mathcal{H},
% \end{align}
% where $\underline{U}_{g}$ and $\overline{U}_{g}$ are the generator minimum and maximum ramp rate limits. Constraint~\eqref{eq:StandardRamp} needs to be modified to consider unit commitments. 

% The standard ramping constraint \eqref{eq:StandardRamp} is modified to 
The auxiliary variable $p^{\text{aux}}_{g, t,\xi}$ is introduced in~\eqref{eq:RampAUX} as equal to the difference in generation of generator $g$ from its minimum output limit ($\underline{U}_{g}$) and zero otherwise. The constraint~\eqref{eq:RampNEW} prevents ramp violations during the startup process. 
\begin{subequations}
\begin{align}
p^{\text{aux}}_{g,t,\xi}&=p_{g,t,\xi}-\overline{p}_{g} z_{g, t}, \quad \forall g \in \mathcal{G}, \: t \in \mathcal{H}, \label{eq:RampAUX} \\
\underline{U}_{g} &\leq p^{\text{aux}}_{g, t+1,\xi}-p^{\text{aux}}_{g,t,\xi} \leq \overline{U}_{g} \quad \forall g \in \mathcal{G},\: t\in \mathcal{H},\: 
\xi \in \Xi .\label{eq:RampNEW}
\end{align}
\end{subequations} 
It is assumed that when a line is switched off, it remains off for the remainder of the day which is enforced by,
\begin{align}
z_{ij,t} \geq z_{ij,t+1} 
\quad \forall \left({i,j} \right) \in \mathcal{L}, t \in \mathcal{H}.
\label{eq:Damaged}
\end{align}
% With constraints \eqref{eq:UC1}-\eqref{eq:UC3}, \eqref{eq:Pg},\eqref{eq:RampAUX}-\eqref{eq:RampNEW}, more realistic generator %and stationary storage energization decisions can be computed in areas prone to wildfires because generators no longer possess the ability to instantaneously ramp-up/down and start-up/shut-down.
To model power flow through the transmission lines, the DC-OPF approximation is used as

\begin{subequations}
\begin{align}
\label{eq:MaxPowerFlow}
p_{ij, t,\xi} &\leq-B_{ij}\left(\theta_{i,t,\xi}-\theta_{j,t,\xi}+\overline{\theta}\left(1-z_{ij,t}\right)\right). \\
\label{eq:MinPowerFlow}
p_{ij,t,\xi} &\geq-B_{ij}(\theta_{i,t,\xi}-\theta_{j,t,\xi}+\underline{\theta}\left(1-z_{ij,t}\right)),\\
\label{eq:ThermalLimit}
\underline{p}_{ij,t} \, z_{ij,t} &\leq p_{ij,t,\xi} \leq \overline{p}_{ij,t}\, z_{ij,t}
\end{align}
\end{subequations}
for all $t\in \mathcal{H}$ $(i,j) \in \mathcal{L}$. Constraints~\eqref{eq:MaxPowerFlow} and~\eqref{eq:MinPowerFlow} limit power flow $p_{ij,t,\xi}$ between maximum and minimum limits, where $B_{ij}$ is the susceptance of the line. When $z_{ij,t} = 1$,~\eqref{eq:MaxPowerFlow} and \eqref{eq:MinPowerFlow} hold with equality. When $z_{ij,t} = 0$, the power flow across transmission line $(i,j)$ at time $t$ is zero and the phase angle $\theta_{i,t,\xi}$ is limited between its maximum ($\overline{\theta}$) and minimum ($\underline{\theta}$) limits. Constraint~\eqref{eq:ThermalLimit} enforces the thermal limits, $\underline{p}_{ij,t}$ and $\overline{p}_{ij,t}$, on power flow, $p_{ij,t,\xi}$. The decision variable, $z_{ij,t}$, is used in constraint \eqref{eq:ThermalLimit} to ensure that if $p_{ij,t,\xi}$ = 0 the line from bus $i$ to $j$ is de-energized. Finally, the bus power balance at each $t\in \mathcal{H}$ and every bus $i \in \mathcal{B}$ is given by the following equality constraint,
\begin{align}
\label{eq:PowerBalance}
\sum_{g \in \mathcal{G}_{i}} p_{g,t,\xi}+\sum_{(i,j) \in \mathcal{L}} p_{ij, t,\xi} -\sum_{d \in \mathcal{D}_{i}} x_{d,t,\xi} p_{d,t,\xi}=0.
\end{align}

% \begin{align}
% \label{eq:PowerBalance}
% &\sum_{g \in \mathcal{G}_i} p_{g,t,\omega}+\sum_{j:i\rightarrow j} p_{ij, t,\omega} -\sum_{d \in \mathcal{D}_i} x_{d,t,\omega} D_{d,t,\omega}=0 \\
% &\hspace{0.5 cm} \forall t \in \mathcal{H} ,\: i \in \mathcal{B}, \: \omega \in \Omega \nonumber
% \end{align}

%To ensure optimal power shutdown, where maximum load delivered increases even with wildfire risk, we  advance our analysis by observing the change in load shedding behavior via the addition of energy storage units. We hope to see that the change in load served is minimized and the wildfire risk incurred at each load is lower than the former case.

\vspace{-2em}
\subsection{Constraint for de-energizing  no less than k lines}
\label{subsection: Constraint for Enacting No Less Than k Damaged Lines}
During the day ahead planning phase, the grid operator has the flexibility to control the maximum number of active lines based on WFPI risk tolerance. Constraint~\eqref{eq:WFPI_Nmk} represents a line shut-off strategy that constrains the total number of active lines so that the total line risk does not exceed $R_{\text{tol}}$ which is an operator-driven parameter (measured in WFPI \cite{WFPI}). $R_{\text{tol}}$ guarantees a certain level of security for the system while also penalizing the operation of transmission lines within regions of higher WFPI, %vegetation flammability levels. 
\begin{align} \sum_{\left(i,j \right) \in \mathcal{L}} z_{ij,t}R_{ij,t} \leq R_{\text{tol}},\quad \forall t \in \mathcal{H} .\label{eq:WFPI_Nmk}\end{align}
% In other words, \eqref{eq:RFire} is the sum of all wildfire risks for energized transmission lines. The line de-energization is a first-stage variable and it is the same across all scenarios by construction. 

Constraint \eqref{eq:Nmk} is similar to the standard $N$-$k$ shut-off strategy in which the operator defines the maximum number of lines that are active in the system for the complete time horizon without considering the wildfire risk~\cite{Bienstock}. 
\begin{align} 
\sum \limits_{\left({i,j} \right) \in \mathcal{L}} z_{ij,t} \leq |\mathcal{L}|-|\mathcal{K}|,\quad \forall t \in \mathcal{H} \label{eq:Nmk}
\end{align}
Operators use~\eqref{eq:WFPI_Nmk} or \eqref{eq:Nmk} to obtain a line shut-off strategy that achieves a lower risk. However, in the case of~\eqref{eq:WFPI_Nmk}, there are multiple solutions with the same number of active lines but with different wildfire risk that still satisfy~\eqref{eq:WFPI_Nmk}. To prioritize a low-risk solution,~\eqref{eq:WFPI_Nmk} is modified to an equality constraint with a slack variable $R_{\text{slack}}$ that minimizes the difference from the minimum wildfire risk for a given number of active lines as
% Constraint \eqref{eq:WFPI_Nmk_slack} is used to balance the minimization of economic costs with tightening the allowable wildfire risk. A simple heuristic is used to bound the slack in the constraint $R^{\max}_{\text{slack}}$ to prevent overusage of the slack variable:
\begin{subequations}
\begin{align} 
\sum \limits_{\left(i,j \right) \in \mathcal{L}} z_{ij,t}R_{ij,t} + R_{\text{slack}}&= R_{\text{tol}},\quad \forall t \in \mathcal{H}, \label{eq:WFPI_Nmk_slack} \\
0 \leq R_{\text{slack}} &\leq \overline{R}_{\text{slack}},
\label{eq:slack_max}\end{align}
\end{subequations}
where $\overline{R}_{\text{slack}}$ is a design parameter chosen to upper bound violation. Total fire risk $ R_{\text{Tot}}$ is defined as the sum of risk $R_{ij,t}$ of all energized transmission lines and is given by,
\begin{align}
    R_{\text{Tot}} =& \sum_{t \in \mathcal{H}} \sum_{(i,j) \in \mathcal{L}}z_{ij,t} R_{ij,t}.
    \label{eq:RFire}
\end{align}
Each risk $R_{ij,t}$ is the WFPI value assigned to the transmission line connecting bus $i$ to bus $j$ at time $t$.
\vspace{-1em}
\subsection{Two Stage Stochastic Problem}
\label{subsection:Two Stage Stochastic Problem}
A two-stage scenario-based formulation is developed in this section to include demand uncertainty. Unit commitments ($z_{g}^{\text{up}}$, $z_{g}^{\text{dn}}$, $z_g$) and transmission line shut-offs ($z_{ij}$) are determined for the entire optimization horizon and thus modeled as first-stage decisions. The generator output ($p_{g,\omega}$) and load served ($x_{d, \omega}$) are determined in the second stage depending on the realized demand ($p_{d,\omega}$). The standard risk-neutral two-stage unit commitment problem optimizes for expected total costs \cite{Noyan}. Wildfires are less frequent high-impact events and call for non-routine operations. A risk-averse decision-making strategy, rather than a purely risk-neutral one, is required to handle the effect of variability in the demand. A natural approach to handling volatility in the costs is a mean-variance approach (e.g. Markowitz approach). However, mean-variance optimization penalizes upper-tail and lower-tail costs equally. Instead, a widely adopted convex and coherent risk metric, conditional value at risk, is used to penalize upper-tail costs \cite{Noyan,Rockafellar, Shaked} resulting in a mean-risk objective function given by,

\begin{align}
&\min_{\bm{z^{\text{up}}_{g}},\bm{z^{\text{dn}}_{g}},\bm{z_{ij}}} \left(1-\beta \right) \mathbb{E}[\Pi_{\bm{\omega}}] + \beta \, \text{CVaR}_\epsilon (\Pi_{\bm{\omega}}), \label{obj:SPSPS}\\
&\text { s.t.} \nonumber \\
&\Pi_{\omega} = f^{\text{uc}}(\bm{z^{\text{up}}_{g}},\bm{z^{\text{dn}}_{g}})+f_{\omega}^{\text{oc}}(\bm{p}_{\bm{g}, \omega}) + f_{\omega}^{\text{VoLL}}(\bm{x}_{\bm{d},\omega},\bm{p}_{\bm{d},\omega})+R_{\text{slack}}, \label{obj:PI_omega}\\
& \text{Line Contingencies: } \{
\eqref{eq:WFPI_Nmk},\eqref{eq:WFPI_Nmk_slack} \& \eqref{eq:slack_max}, \, \text{or} \,\eqref{eq:Nmk} \} \, \text{and} 
 \, \eqref{eq:Damaged}, \nonumber \\ %\eqref{eq:zixd}-
& \text{Unit Commitment Constraints: }
\, \eqref{eq:UC1}-\eqref{eq:UC3}, \nonumber \\
&\text{CVaR Constraints: } \, \eqref{eq:CVARcon1}-\eqref{eq:CVARcon3}, \nonumber \\
&\text{Generator Capacity Bounds: } 
 \,\eqref{eq:Pg}, \nonumber \\
& \text{Generator Ramping Constraints: } 
 \, \eqref{eq:RampAUX}-\eqref{eq:RampNEW}, \nonumber\\
& \text{Optimal Power Flow Constraints: } \, \eqref{eq:MaxPowerFlow}-\eqref{eq:ThermalLimit}, \nonumber\\
&\text{Day-ahead Demand Balance Constraints: } \, \eqref{eq:PowerBalance}, \nonumber \\
&\forall t \in \mathcal{H} ,\: i \in \mathcal{B} ,\: \forall \omega \in \Omega. \nonumber
\end{align}
In~\eqref{obj:SPSPS}, the scenarios are drawn from the historical data by using a scenario reduction technique which selects a set of $K$ demand scenarios, $\omega \in \Omega = \{\omega_1, \dots, \omega_K \}$. 
% A copy of decision variables is also created for each scenario indexed by $\omega$ in the following sections.\
% A scenario approximation is used to approximate the expected value stochastic objective: $\min \left(1-\beta \right) \mathbb{E} \left[ \text{\Pi }_{\bm{\xi}}(\cdot,\bm{\xi}) \right] + \beta \, \text{CVaR}_\epsilon(\text{\Pi }_{\bm{\xi}}(\cdot,\bm{\xi}))$. 
% The scenario approximated first stage optimization is then given by, 
Furthermore, the outer weighted sum with weighting parameter $\beta$ is the balance between the expected value of the total costs for each scenario (denoted as $\Pi_{\omega}$) among all scenarios $\omega \in \Omega$, and its $\epsilon$-quantile expected shortfall or  $\text{CVaR}_{\epsilon}(\Pi_{\omega})$. 
% The relationships between total economic costs, $\Pi_{\omega}$, for each scenario, $\omega$, and along with the probability of that scenario $\pi_{\omega}$ change depending on the definitions of the risk measure. A scenario's economic cost is a sum of the unit commitment costs, operating costs, and the value of lost load costs. } 
% It has been shown in \cite{Shaked, Royset, Rockafellar} and other related works that mean-CVaR optimization is consistent with second-order stochastic dominance 
% (if every realization of asset of $Y_2$ is not smaller than $Y_1$ ($Y_1\leq Y_2$) then the risk of $Y_2$ is not less than the risk of $Y_1$ ($\mathcal{R}(Y_1)\leq \mathcal{R}(Y_2)$) 
% and convex preserving in all $0\leq \beta \leq 1$-- making $\text{CVaR}_{\epsilon}$ a suitable risk metric for unit commitment analysis during wildfire season.
% For random variables with continuous distributions, $\text{CVaR}_{\epsilon}$ is often described as the expected shortfall of costs above a certain confidence level $\epsilon$.
It can be shown that $\text{CVaR}_\epsilon$ is translationally invariant \cite{Noyan}, that is,
\begin{align*}
    \text{CVaR}_\epsilon ( \Pi_\omega)=f^{\text{uc}}(\bm{z^{\text{up}}_{g}},\bm{z^{\text{dn}}_{g}})+\text{CVaR}_\epsilon(\bm{\Phi_{\omega}})+ R_{\text{slack}},
\end{align*}
where $\bm{\Phi_\omega}$ is defined as,
\begin{align}
    \bm{\Phi_\omega} = f_{\omega}^{\text{oc}}(\bm{p}_{\bm{g}, \omega}) + f_{\omega}^{\text{VoLL}}(\bm{x}_{\bm{d},\omega},\bm{p}_{\bm{d},\omega}).
    \label{eq:Phi_omega}
\end{align}
% $\text{CVaR}_\epsilon\biggl(f^{\text{uc}}(\bm{z^{\text{up}}_{g}},\bm{z^{\text{dn}}_{g}})+ R_{\text{slack}}+\bm{\Phi_{\omega}}\biggr)=\text{CVaR}_\epsilon(\bm{\Phi_{\omega}})+f^{\text{uc}}(\bm{z^{\text{up}}_{g}},\bm{z^{\text{dn}}_{g}})+ R_{\text{slack}}$
Therefore, the objective function for the two-stage problem that considers expected and the conditional value at risk ($\text{CVaR}_{\epsilon}$) of total costs, $\Pi_{\omega}$, across scenarios can be split into first-stage costs ($f^{\text{uc}}(\bm{z^{\text{up}}_{g}},\bm{z^{\text{dn}}_{g}})+ R_{\text{slack}}$) and the mean-CVaR second stage costs ($\bm{\Phi}$) that are described next.
% or $\left(1-\beta \right) \mathbb{E}[\bm{\Phi_{\omega}}] + \beta \text{CVaR}_\epsilon (\bm{\Phi_{\omega}})$.

\subsubsection{First Stage Formulation}
\label{subsubsection:First Stage Formulation}
In the first stage of the two-stage unit commitment problem, the operator must make deterministic start-up, shut-down, and de-energization decisions in order to fulfill uncertain demand: %with probability one:
% $\bm{z^{\text{up}}_{g}},\bm{z^{\text{dn}}_{g}},\bm{z_{ij}}$
\begin{align}
&\min_{\bm{z^{\text{up}}_{g}},\bm{z^{\text{dn}}_{g}},\bm{z_{ij}}} \qquad f^{\text{uc}}(\bm{z^{\text{up}}_{g}},\bm{z^{\text{dn}}_{g}})+\left(1-\beta \right) \mathbb{E}[\bm{\Phi_{\omega}}] + \nonumber\\
&\qquad \qquad \qquad\beta \, \text{CVaR}_\epsilon (\bm{\Phi_{\omega}}) +R_{\text{slack}} \label{obj:SPSPS1}\\
&\text { s.t.} \nonumber \\
& \text{Line Contingencies: } \eqref{eq:Damaged}, \, \text{and} \,  \{
\eqref{eq:WFPI_Nmk}, \, \text{or}\, \eqref{eq:Nmk},\, \text{or}\, \eqref{eq:WFPI_Nmk_slack}-\eqref{eq:slack_max} \} \,  \nonumber \\ %\eqref{eq:zixd}-
& \text{Unit Commitment Constraints: }
\, \eqref{eq:UC1}-\eqref{eq:UC3} \nonumber \\
&\text{CVaR Constraints: } \, \eqref{eq:CVARcon1}-\eqref{eq:CVARcon3} \nonumber 
\end{align}

\subsubsection{Second Stage Formulation}
\label{subsubsection:Second Stage Formulation}
In the second stage, 
% of the two-stage unit commitment problem, 
the operator solves an economic dispatch problem for each scenario:
% $\omega \in \Omega$:
% &\Phi_{\omega} (\bm{z^{\text{up}}_{g}},\bm{z^{\text{dn}}_{g}},\bm{z_{ij}},\bm{p}_{\bm{d},\omega}) =\nonumber \\
% &\min_{\bm{p}_{\bm{g},\omega},\bm{x}_{\bm{d},\omega}} f_{\omega}^{\text{oc}}(\bm{p}_{\bm{g}, \omega}) + f_{\omega}^{\text{VoLL}}(\bm{x}_{\bm{d},\omega},\bm{p}_{\bm{d},\omega}), \label{obj:SPSPS2}\\
\begin{align}
& \min_{\bm{p}_{\bm{g},\omega},\bm{x}_{\bm{d},\omega}} \quad \bm{\Phi_{\omega}}\label{obj:SPSPS2}\\
&\text { s.t.} \nonumber \\
& \text{Generator Capacity Bounds: } 
 \,\eqref{eq:Pg} \nonumber \\
&\text{Generator Ramping Constraints: } 
 \, \eqref{eq:RampAUX}-\eqref{eq:RampNEW} \nonumber\\
&\text{Optimal Power Flow Constraints: } \, \eqref{eq:MaxPowerFlow}-\eqref{eq:ThermalLimit} \nonumber\\
&\text{Day-ahead Demand Balance Constraints: } \, \eqref{eq:PowerBalance} \nonumber \\
%\eqref{eq:max charge rate}-\eqref{eq:SOC start end},
&\quad \forall t \in \mathcal{H} ,\: i \in \mathcal{B} ,\: \forall \omega \in \Omega \nonumber
\end{align}
where, $\bm{z^{\text{up}}_{g}},\bm{z^{\text{dn}}_{g}},\bm{z_{ij}}$, are obtained from the first stage.

\subsubsection{Scenario-Based Discretization}
% of Mean-CVaR Second Stage Objective
\label{subsubsection:Scenario-Based Discretization of Mean-CVaR Second Stage Objective}

To reduce the computational complexity of taking the expectation over all possible demand scenarios, a tree scenario reduction algorithm, developed in \cite{Growe-Kuska}, is used to reduce the number of demand scenarios to a smaller finite set of demand scenarios $\Omega$ from the day-ahead forecast. The probability of occurrence of a given scenario $\omega \in \Omega$ is denoted by $\pi_{\omega}$. The expectation is then taken over the set of reduced scenarios resulting in,
\begin{align}
    \mathbb{E}[\bm{\Phi_{\omega}}] = \sum_{\omega \in \Omega} \pi_{\omega} \bm{\Phi_{\omega}}.
\end{align}
%In this work, the tree scenario reduction takes three months' demand forecasts for the total of the three areas of the RTS-GMLC as an input and outputs the 5 most likely area demand scenarios for the upcoming day. That system-wide RTS GMLC demand is then divided among each node in proportion to the static demands given in the IEEE 14-bus bus data set. 
For a discrete probability distribution, $\text{CVaR}_{\epsilon}$ of $\bm{\Phi_{\omega}}$ is mathematically defined as,
\begin{align} 
\text{CVaR}_\epsilon (\bm{\Phi_{\omega}}) = {\min_{\nu} } \left\lbrace { \nu + \frac{1}{1-\epsilon } {\mathbb {E}} } \left[ {\max } \left\lbrace {\bm{\Phi_{\omega}}}-\nu, 0 \right\rbrace \right] \right\rbrace\!.
\label{eq:CVARdef}
\end{align}
At optimality, the auxiliary variable, $\nu$, is commonly referred to as the value at risk (VaR) or minimum cost at an $\epsilon$-confidence level. 
% VaR is not used to represent the risk in the mean-risk objective because its inclusion would make the objective non-convex \cite{Royset}. 
% Intuitively, $\text{CVaR}_{\epsilon}$ can be thought of as a two-stage decision in which an operator selects the lowest possible economic cost $\nu$ today that would later generate the smallest expected shortfall above a given $\epsilon$-quantile between uncertain future economic costs $\Pi_{\omega}$ and the present cost, $\nu$ \cite{Royset}. 
The $\text{CVaR}_{\epsilon}(\bm{\Phi_{\omega}})$ can be approximated as a linear program with $\gamma_{\omega}:={\max } \left\lbrace  \Phi_{\omega }- \nu, 0 \right\rbrace$, $\nu$ as an auxiliary variable, and $\epsilon \in \left[0,1\right]$.
\begin{subequations}
\begin{align}
& \nu + \frac{1}{1-\epsilon} \sum\limits _{\omega \in \Omega } \pi _\omega \gamma _\omega = \text{CVaR}_\epsilon(\bm{\Phi_{\omega}}) %&\, \forall \, \omega \in \Omega 
\label{eq:CVARcon1} \\ &  \Phi_\omega-\nu \leq \gamma _\omega &\,\forall \, \omega \in \Omega \label{eq:CVARcon2}\\ & 0 \leq \gamma _\omega &\, \forall \, \omega \in \Omega \label{eq:CVARcon3}
\end{align}
\end{subequations}

Two parameters contribute to the operator's level of risk-averseness: $\epsilon$ and $\beta$. By definition, $\epsilon$ adjusts the operator's averseness to high-tail economic costs. The economic costs in each scenario are strongly influenced by the amount of demand served in each scenario due to the high value assigned to the Value of Lost Load. When $\epsilon=0$, $\text{CVaR}_{\epsilon}(\mathrm{\Phi }_{\bm{\omega}})=\mathbb{E}(\mathrm{\Phi }_{\bm{\omega}})$ and the operator is risk-neutral. When $\epsilon=1$, $\text{CVaR}_{\epsilon}(\mathrm{\Phi }_{\bm{\omega}})=\sup\{\mathrm{\Phi }_{\bm{\omega}}\}$ and the operator is the most-risk averse. The confidence level, $\epsilon$, is fixed at 95\% in this work. For normally distributed cost scenarios, 95\% is a commonly used baseline confidence level equivalent to considering scenarios over 1.96 standard deviations from the mean. By construction, increasing $\beta$ (for a fixed $\epsilon$) gives more penalty to the more risk-averse term in the objective $\text{CVaR}_{\epsilon}$ and increases the mean-$\text{CVaR}_{\epsilon}$ objective. Due to the shifting of weight from mean to $\text{CVaR}_{\epsilon}$, increasing $\beta$ (for a fixed $\epsilon$) increases the mean costs and decreases the $\text{CVaR}_{\epsilon}$ costs in the mean-$\text{CVaR}_{\epsilon}$ objective. 

\subsubsection{Real-Time Optimization Test}
\label{subsubsection:Real-Time Optimization Test}

If the operator would like to test the performance of the day-ahead generator commitments and transmission line shut-off decisions on real-time demand, the system operator can implement the first-stage decisions and redispatch the first-stage decisions to avoid infeasibility. 
% These decisions are influenced by the demand realized at each bus.
% $d \in \mathcal{D}$ of the demand from a finite, tree-reduced set of demand scenarios with improved forecasts given real-time data,  $\omega' \in \Omega^{\text{RT}}$. %, and demand-specific wildfire risk values, $R_d$. 
The real-time test can implemented in a receding horizon fashion \cite{Rawlings} to simulate adjustments to the real-time demand. 
% At time $t$, the optimization problem is solved for the horizon of length $H$ and only the first control action is implemented. 
% Then, all time-coupled variables such as the generator setpoints are advanced one time step into the future and the incoming real-time demand operational costs at $t+1$ are recorded. 
% The optimization is solved again over a prediction window of the same length as before. The optimization is repeated for all times steps $t$ in the optimization horizon $\mathcal{H}$. 
Instead of optimizing costs over forecasts of demand scenarios $\omega \in \Omega$, realized samples of demand, $p_{d,\omega'}$, are used from a set of real-time samples  $\Omega^{\text{RT}}$. %One realization of demand is used in the analysis in Section~\ref{subsection:Results for the SPSPS}. %In general, these realizations could also be drawn from a set of improved real-time forecast scenarios $\omega'\in \Omega^{\text{RT}}$.
% \begin{align}
% \min  \: &\sum_{\omega'\in \Omega^{\text{RT}}} \pi_{\omega'} {\Pi }_{\omega' } %+ C^{\text{OC}}_{\tau}
% \label{obj:SPSPSRT}\\
% \text { s.t. } %&\eqref{eq:zixd},
% % &\eqref{eq:Pg},\eqref{eq:RampAUX}-\eqref{eq:RampNEW}, \eqref{eq:MaxPowerFlow}-\eqref{eq:ThermalLimit}%,\eqref{eq:max charge rate}-\eqref{eq:SOC start end} 
% \nonumber\\
% % & \hspace{-0.85 cm} \text{\Pi }_{\omega'} = \sum\limits_{t \in \mathcal{H}_p(\tau)}  \left(\mathop \sum \limits_{d \in \mathcal{{D}}} \text{VoLL}_d\left(1-x_{d,t,\omega'}\right)D_{d,t,\omega'} + \mathop \sum \limits_{g \in \mathcal{G}} {c_g p_{g,t,\omega'}} \right) %\:\forall \, \omega' \in \Omega^{\text{RT}} 
% % \\
% & \text{Generator Capacity Bounds: } 
%  \,\eqref{eq:Pg} \nonumber \\
% &\text{Generator Ramping Constraints: } 
%  \, \eqref{eq:RampAUX}-\eqref{eq:RampNEW} \nonumber\\
% &\text{Optimal Power Flow Constraints: } \, \eqref{eq:MaxPowerFlow}-\eqref{eq:ThermalLimit} \nonumber\\
% &\text{Real-time Demand Balance Constraints: } \, \eqref{eq:PowerBalance} \nonumber \\
% %\eqref{eq:max charge rate}-\eqref{eq:SOC start end},
% % &\sum_{g \in \mathcal{G}_i} p_{g,t}+\sum_{j:i\rightarrow j} p_{ij, t}%+\sum_{s \in \mathcal{S}_i} p^{\text{dis}}_{s,t} -\sum_{s \in \mathcal{S}_i}p^{\text{char}}_{s,t}
% % %\nonumber\\ \label{eq:PowerBalanceS2}
% % -\sum_{d \in \mathcal{D}_i} x_{d,t} D_{d,t,\omega'}= 0, \\
% & \hspace{0.5 cm} \:\; i \in \mathcal{B} ,\: \forall \omega' \in \Omega^{\text{RT}}, \: \forall t \in \mathcal{H}_p(\tau) \nonumber \\
% & \text{then repeat optimization } \: \forall \tau \in \mathcal{H} \nonumber
% \end{align}

\subsubsection{Calculation of Expected Demand and Production Costs}
\label{subsubsection:Calculation of Expected Demand and Production Costs}
In Section \ref{section:Results}, we break down the total economic costs into its cost components: commitment costs, operating costs, the VoLL, and wildfire risk slack $R_{\text{slack}}$. Since operating costs and VoLL costs are given for each scenario, $\omega$, \eqref{eq:ExpDemand} and \eqref{eq:ExpProdCost} are used to calculate the expected demand (which is proportional to VoLL) and expected production costs resulting from the second-stage decisions. The expected total demand served \eqref{eq:ExpDemand} is defined as,
\begin{equation}
    \mathbb{E}[\bm{D_{\text{Tot},\omega}}] = \sum_{\omega\in \Omega} \pi_{\omega} \left(\sum_{t\in \mathcal{H}}\sum_{d \in \mathcal{D}} x_{d,t,\omega}p_{d,t, \omega}\right),
    \label{eq:ExpDemand}
\end{equation}
where, $p_{d,t,\omega}$ is the predicted demand at bus $d$ and at time $t$ for each scenario $\omega$, and $x_{d,t,\omega}$ is the proportion of that total demand $p_{d,t,\omega}$ served. $\pi_{\omega}$ is the probability of the demand scenario $\omega$.
% In the deterministic case, $p_{d,t}(\bar{\xi})$ is the mean predicted demand. 
% $\pi_{\omega}$ is the probability of a given scenario $\omega \in \Omega$.
Similarly, the expected operating cost is obtained as the weighted sum of $f_\omega^{\text{oc}}(\bm{p}_{\bm{g},\omega})$ over $\omega\in\Omega$, 
% In Section \ref{section:Results}, we only consider variable generator production costs in the operating cost objective.
% variable production cost is defined by \eqref{eq:ExpProdCost} as the sum of every active generator's $g$ production $p_{g,t}$ times its variable cost of generation $c_g$ for every time step in the optimization horizon. 
\begin{align}
\mathbb{E}[\bm{\text{Op. Cost}_{\omega}}]=\sum_{\omega\in \Omega} \pi_{\omega} f_{\omega}^{\text{oc}}(\bm{p_{g,\omega}})
\label{eq:ExpProdCost}
\end{align}
%In Section \ref{section:Results}, we only consider variable generator production costs in the operating cost objective.

%\subsubsection{Remark}
%\begin{remark}
% The deterministic PSPS formulation is a special case of the stochastic PSPS problem in which it is assumed that the uncertainty is captured in a single scenario that represents the expected demand (i.e. $\mathbb{E}[\bm{p_{d,\xi}}]$). 
%\end{remark}

\section{Results}
\label{section:Results}
In this Section, the differences in optimal decisions and costs for the deterministic and stochastic PSPS frameworks are analyzed for day-ahead unit commitment on the IEEE 14-bus system and for the IEEE RTS 24-bus system.

\subsection{Data and Test Case Description} \label{subsection:Data and Test Case Description}

WFPI has been used in recent research works, e.g.~\cite{Rhodes, Kody, Rhodes2, Kody2}, to assess the risk of damage to power system components from nearby wildfires. WFPI describes the ratio of live to dead fuel and includes modifiers for wind speed, dry bulb temperature, and rainfall~\cite{WFPI}. 
% WFPI values range from $0$ to $150$. Values $249-254$ mark land outside the United States, ice, agricultural land, barren land, marshland, and bodies of water; all of the terrain conditions with WFPI values from $249-254$ are considered to have 0 WFPI. In~\cite{Rhodes, Kody, Rhodes2, Kody2}, day-ahead forecasts of WFPI are taken directly from the USGS database. 
The USGS generates WFPI forecasts for the continental US up to seven days in the future. 
% \textcolor{blue}{WFPI values assigned to the IEEE 14 bus network is depicted in Figure~\ref{fig:IEEE14colored} and WFPI values assigned to the IEEE RTS 24 bus system is depicted in Figure~\ref{fig:IEEE24colored}.}%Specifically, the region under consideration is Southern Nevada, and Northwestern Arizona.
\subsubsection{Grid models description}
\label{subsubsection:Grid models description}
\begin{figure}[ht]
\vspace{-2em}
\centering
\includegraphics[width=\columnwidth]{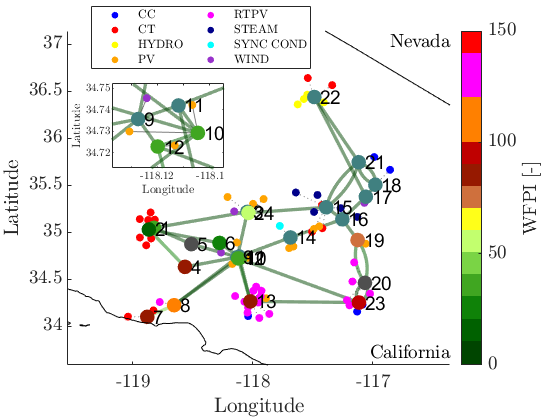}
\vspace{-2em}
\caption{IEEE RTS 24-bus system schematic with each transmission line and bus highlighted to depict its WFPI for July 31, 2020. The geographic layout is obtained from \cite{RTSGMLC}. Zero-risk buses are 3, 5, 9, 11, 14-18, 20-22 and zero-risk lines are 14, 23-26, 28, 30-33, 38. 
% There are four line pairs:  25/26, 32/33, 33/34, and 36/37. Each line in a pairing is assumed to have the same risk.
}
\label{fig:IEEE24colored}
\end{figure}

The grid models used for analyses are the modified IEEE 14 bus system as shown Fig.~\ref{fig:IEEE14colored} and the IEEE RTS 24-bus system as shown in Fig.~\ref{fig:IEEE24colored} along with their assigned WFPI values. 

The IEEE 14 bus system consists of 2 generators (at buses 1 and 2), 11 loads (one load each at buses 2-6 and 9-14), and 20 transmission lines. The buses are assumed to be located at 14 locations in Southern Nevada and Northwestern Arizona that typically experience large wildfire risk throughout the year. Generator costs, power, and energy capacities are provided in~Table~\ref{table:IEEE14GenStats}. Hourly load profiles are obtained from~\cite{RTSGMLC}. The load profile for each of the 14 buses is then obtained by scaling the profile so that the maximum load is equal to the load defined in~\cite{PowerGridLibIEEE14}.

The IEEE RTS 24 bus system consists of 68 generating sources (3 wind generators, 20 roof-top PV (RTPV) clusters, 14 utility-scale PV units, 4 hydro, 1 synchronous condenser, 5 combined cycle (CC) generators, 15 combustion turbine (CT), 6 steam-powered generators), 17 loads, and 38 transmission lines. More information about the location, capacity, and costs of operation of each generator type are provided in~\cite{RTSGMLC}. The model is located within a region of Southern California near Los Angeles. The IEEE RTS 24 bus system is equivalent to the northwest region (i.e. Region 300) of the 73-bus RTS-GMLC~\cite{RTSGMLC}. The daily regional load profile given in Region 300 of the RTS GMLC system datasheet is projected onto a max-scaled version of the daily load experienced at each of the 24 bus locations \cite{RTSGMLC}. Hourly load profiles are obtained from~\cite{RTSGMLC}.

\subsubsection{Transmission line WFPI assignment}
\label{subsubsection:Transmission line WFPI assignment}
\begin{figure}[ht]
\centering
\vspace{-1em}
\includegraphics[width=0.95\columnwidth]{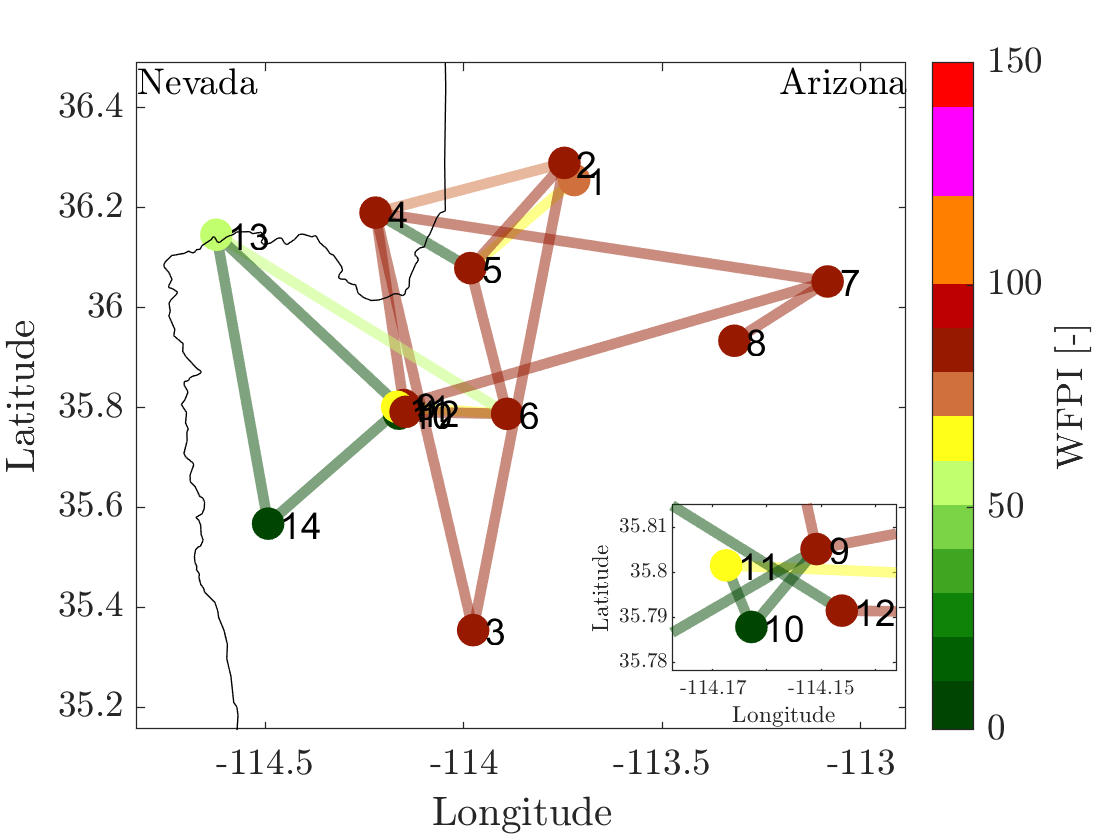}
\vspace{-1em}
\caption{IEEE 14-bus system schematic with each transmission line and bus color-coded to depict its wildfire risk value. %Note the arbitrary layout of the buses comes from \protect\cite{PowerGridLibIEEE14}; the distances between buses are not to scale % according to \cite{PSCAD}. 
WFPI values were recorded on October 11th, 2015 in the southwestern U.S.  \protect\cite{WFPI}. Buses are located in Southern Nevada and Northwestern Arizona.}
\label{fig:IEEE14colored}
%data: worskpace_econWFPI101122IEEE14Bcases 09_Feb_2024_20_21_12.mat
%script:fig2IEEE14colored.m
\end{figure}

From the USGS fire data products page \cite{WFPI}, day-ahead forecasts are downloaded in .tiff format and combined to form timeseries of WFPI predictions for each of the 14 bus locations in the IEEE 14-bus system. Each .tiff file contains a mesh in which each point is associated with a WFPI value and a pixel with $(x,y)$ coordinates. %Pixel coordinates are translated into longitude and latitudes. % Since the precision of the bus longitude and latitude is finer than the precision of the .tiff file, 
The four nearest neighbors from the .tiff grid mesh are averaged to form the WFPI forecast of the bus location. Forecasts for transmission lines are assumed to be the maximum WFPI value experienced on the straight line between connecting buses. In this way, bus WFPI forecasts provide a conservative estimate of the transmission line WFPIs. The bus and transmission line longitude/latitude are shown in~Fig.~\ref{fig:IEEE14colored}. The bus longitude/latitude coordinates for the buses on the IEEE RTS 24-bus system are predefined in \cite{RTSGMLC}. We assume the transmission lines exist as presented by \cite{RTSGMLC}. A collection of WFPI values is assigned to each transmission line, $(i, j)$, via the Bresenham algorithm \cite{Bresenham} and the buses at the ends of the lines. The WFPI value assigned to the line in the optimization is the average WFPI value in each collection.

\begin{table}[ht]
\begin{center}
\caption{IEEE 14-bus system generator statistics \protect\cite{RTSGMLC}, \protect\cite{PowerGridLibIEEE14}}
\begin{tabular}{l 
>{\raggedleft\arraybackslash}m{1.2cm} 
>{\raggedleft\arraybackslash}m{1.8cm} 
>{\raggedleft\arraybackslash}m{1.6cm}}
\hline
& $\overline{p}_g$, $\underline{p}_g$ (MW) & $\overline{U}_g$, $\underline{U}_g$ (MW/h)& $t_g^{\text{MinUp}}, t_g^{\text{MinDn}}$ (h) \\ \hline
Gen 1 & 340, 0 & $\pm$ 248.4 & 8.0, 4.0\\
Gen 2 & 59, 0 & $\pm$ 22.0 & 2.0, 2.0\\
\hline & $C_g$ (\$/MWh)& $C^{\text{up}}_g, C^{\text{dn}}_g$ 
  (\$)%$\times 10^3$) 
  &\\
\hline
Gen 1 & 7.92 & 280, 280 %28.0 28.0 
&  \\
Gen 2 & 23.27 & 56, 56 %5.6, 5.6 
& \\
\hline
\end{tabular}
\label{table:IEEE14GenStats}
% FYI Actual Mapping of IEEE Lines to RTS-GMLC Lines [42,44,43,45,46,48,50,51,49,47,52,53,54,55,56,58,59,57,61,60]
% Variables of Interest
%P_gmax
%P_gmin
%R_gmax=R_gmin
%Tmin_shut
%Tmin_start
%Gen_cost
%Gen_start_cost
%Gen_shut_cost
\end{center}

\end{table}

% Additional simulations are performed on the Reliability Test System Grid Modernization Lab Consortium's (RTS-GMLC) 73-bus case \cite{RTSGMLC}, which is representative of a portion of Southern California's transmission grid (see Figure \ref{fig:GMLCcolored}). The network contains 213 distributed generators (combustion turbine, steam, combined cycle, synchronous condensers, hydropower, wind, concentrated solar power, roof-top PV, commercial PV, and nuclear), 51 loads, and 120 transmission lines. The mapping of USGS WFPI %and WLFP maps 
% to the buses and transmission lines on the RTS GMLC can be seen in Figure \ref{fig:GMLCcolored}.

\begin{figure}[b]
\centerline{\includegraphics[width=0.9\columnwidth]{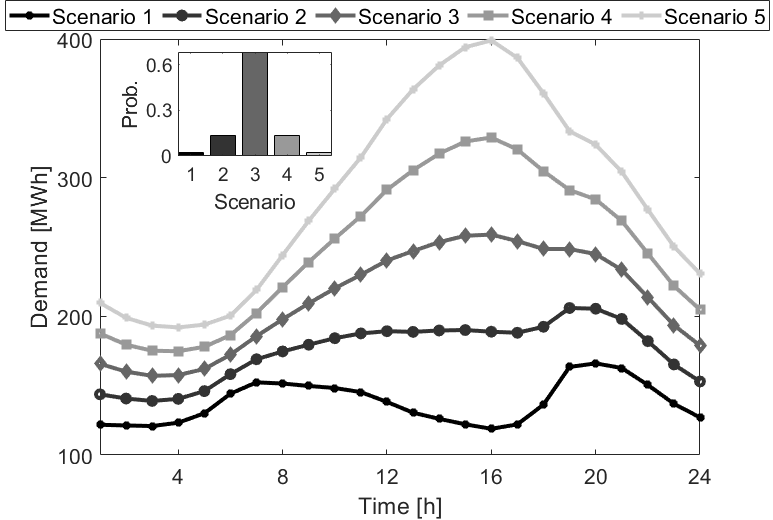}}
\caption{Load scenarios (mean, $\pm$ 1 std. dev. and $\pm$ 2 std. dev.) for the IEEE 14-bus system derived from the tree reduction load profiles for the RTS-GMLC for October 11, 2020. The probabilities of occurrence of each demand scenario are shown in the upper left corner.}
\label{fig:Load_Profiles_IEEE14}
%data: worskpace_econWFPI101122IEEE14Bcases 09_Feb_2024_20_21_12.mat
%script:fig3Load_Profiles_IEEE14.m
\end{figure}

%\subsubsection{Translating Regional Demand Forecasts from RTS_GMLC to Demand Scenarios for the IEEE 14-bus}
To implement the two-stage stochastic PSPS, demand scenarios are generated via a tree reduction algorithm from \cite{Growe-Kuska}. The optimization horizon of interest is one day. %; however, an extra day is needed so that the receding horizon control window can remain the same length throughout the optimization horizon. 
The tree scenario reduction method analyzes three months of prior demand forecasts for the total cumulative load for the region in which the 14 IEEE buses reside. The output is a scenario tree of 5 representative demand scenarios for the next day. 
The five demand scenarios % generated from the tree reduction algorithm, depicted in \ref{fig:Load_Profiles}, 
are max-scaled and proportionally mapped to each of the static loads given in the IEEE 14-bus system datasheet. To adjust the scenario set to include less probable high-demand scenarios, at each time step a normal distribution is fit to the five demand scenarios.
% the distribution of the five scenarios is then mapped to a normal distribution whose mean and standard deviation at each time point are the sample means and standard deviations of the tree reduction algorithm load at the same time point. 
At each time step, $t\in \mathcal{H}$, there are five new samples:  the sample mean, sample mean $\pm$ 1 standard deviation, and sample mean $\pm$ 2 standard deviations for that given time step. The five new scenarios include timeseries of all sample mean, all sample means $\pm$ 1 standard deviation, and sample mean $\pm$ 2 standard deviations. The five newly generated scenarios are shown in Figure~\ref{fig:Load_Profiles_IEEE14}. The stochastic optimization problem is converted to a deterministic problem with one power balance constraint for each demand scenario \eqref{eq:PowerBalance}. 
\vspace{-1em}

\subsection{Comparison of placement of wildfire risk in objective or as constraint}
\label{subsection:Comparison of placement of wildfire risk in objective or as constraint}
The OPS model in \cite{Rhodes, Kody, Rhodes2, Kody2} uses a weighted objective function that trades off maximizing demand served with minimizing wildfire risk. Instead, we capture the wildfire risk in a constraint rather than a secondary objective. 
% This method is called as the $\epsilon$-constraint method for bi-objective optimization problems \cite{Haimes}. 
In this section, we compare the deterministic version of our Two-Stage Stochastic Public Safety Power Shut-off (SPSPS) to the OPS model from \cite{Rhodes, Kody, Rhodes2, Kody2}.

The drawback of the OPS model's weighted objective function philosophy is that the weighted OPS objective does not capture all Pareto optimal points corresponding to optimal wildfire risk and optimal operating costs. Whereas our SPSPS model is able to capture all the Pareto optimal points with careful selection of $R_{\text{tol}}$. Consider the following OPS objective function which is a balance between operating costs and risk,
% We modify the OPS problem to be a balance between operating costs and risk. The modified OPS that includes operating costs and a VoLL cost for total demand is the following:
% \textcolor{blue}{
\begin{align}
    \text{min}\: &\left(\frac{(1-\alpha)}{\overline{\Pi}^{NoUC}_{\text{det}}}\Pi^{NoUC}_{\text{det}} + \frac{\alpha}{\overline{R}_{\text{Tot}}} R_{\text{Tot}}\right),
    \label{eq:OPSform}
\end{align}
% The total wildfire risk, $R_{\text{tot}}=\sum^H_{t=1}\sum_{(i,j) \in \mathcal{L}}R_{ij,t}z_{ij,t}$ is the sum of the risk associated with energized components. 
where $R_{\text{Tot}}$ is defined in~\eqref{eq:RFire}, and $\Pi^{NoUC}_{\text{det}}$ only includes operating costs and VoLL, 
\begin{align}
    \Pi^{NoUC}_{\text{det}} = f^{\text{oc}}(\bm{p_{g}}) + f^{\text{VoLL}}(\bm{x_d},\mathbb{E}[\bm{p_{d,\xi}}]).
\end{align}
Each objective in \ref{eq:OPSform} is normalized by its maximum value;  $\overline{\Pi}^{NoUC}_{\text{det}}$ is the maximum achievable operating cost and $\overline{R}_{\text{Tot}}$ the maximum wildfire risk. The total wildfire risk $R_{\text{Tot}}$ in the objective function~\eqref{eq:OPSform} is non-convex due to the binary de-energization decisions $z_{ij,t}$ resulting in non-convex multi-objective feasible space. Instead, the deterministic SPSPS model without unit commitment costs and constraints is,
\begin{align}
    &\min \qquad \Pi^{NoUC}_{\text{det}},\label{eq:SPSPSform}\\
&\text{s.t} \qquad R_{\text{Tot}} \leq R_{\text{tol}}. \nonumber
\end{align}
% Even if we modify the OPS objective to become a weighted balance between minimizing operating costs and minimizing wildfire risks, there is still a notable  The OPS problem contains an objective function that is a balance between demand and risk. 
% the OPS's weighted sum of objective functions is not guaranteed to capture all Pareto optimal points.

\begin{figure}[ht]
\centerline{\includegraphics[width=\columnwidth]{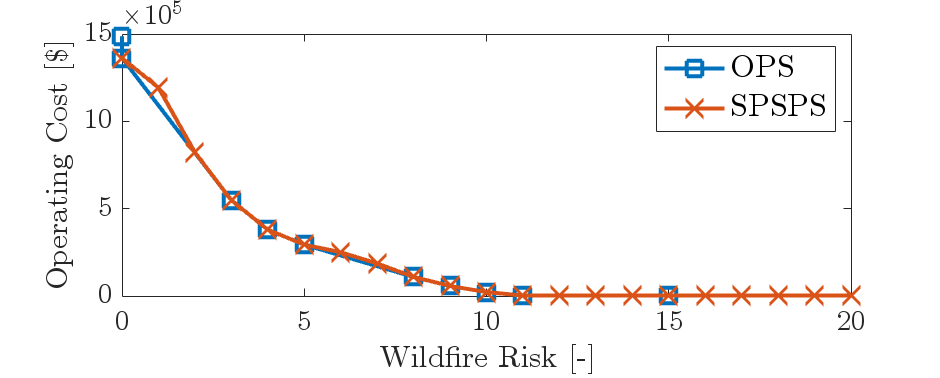}}
\caption{Plots depicting the Pareto Optimal Operating Cost and Optimal Wildfire Risk from the OPS and SPSPS models assuming $R_{ij,t}=1$ optimized for one time step. The OPS model fails to capture Pareto Optimal points at $R_{tot}=1,6, \text{and } 7$.}
\label{fig:OPSvSPSPS}
\end{figure}

In~\eqref{eq:SPSPSform}, the wildfire risk is added to the set of constraints rather than to the objective function. This reformulation is commonly referred to as the $\epsilon$-constraint method. \cite{Haimes} shows that this method can be applied to both non-convex and convex multi-objective feasible spaces to capture all Pareto Optimal points.
In Figure \ref{fig:OPSvSPSPS}, we plot the Pareto curves of optimal operating costs (only generator production and VoLL costs) versus total wildfire risk for the first hour of the day for the OPS and SPSPS models for the IEEE 14-bus system. In order to demonstrate that the SPSPS objective can capture all Pareto optimal solutions, we assume all lines have the same risk equal to 1 $R_{i,j}=1$, and $R_{\text{tol}}$ is swept from 0 (no active lines) to 20 (all lines are active). This makes sure that there is a Pareto optimal solution for all possible number of active lines. For general bi-objective problems, \cite{Haimes} recommends an equi-distant bisection method for $R_{\text{tol}}$. Selecting an appropriate $R_{\text{tol}}$ step size is important for obtaining the entire optimal Pareto front and depends upon the system under consideration which will be explored in future work.
The OPS objective weight $\alpha$, balancing cost and risk is swept from 0 to 1 with a step size equal to 0.01.
% We do not consider unit commitment constraints or commitment costs. We use the deterministic version of the SPSPS; the OPS is already deterministic. 
In Figure \ref{fig:OPSvSPSPS}, it can be seen that the Pareto Optimal points corresponding to 1, 6, and, 7 wildfire risk, are not captured by the OPS model. Whereas, the SPSPS model is able to capture all Pareto optimal points. In this special case when all $R_{ij,t}$ are equal, the wildfire risk can be easily mapped to the number of active lines, making the wildfire risk tolerance more understandable than finding an appropriate range for alpha to sweep. In the following sections, wildfire risk obtained from USGS is used as described in the previous section. In general, SPSPS gives the operator a wider range of possible optimal solutions.
\vspace{-1em}
\subsection{Comparison of line outage strategies (IEEE 14-bus)}
\label{subsection:Comparison of Different Line Outage Strategies}
Five deterministic line outage strategies are shown in~Table~\ref{table:Benchmark}. For simplification, only the load
% the comparisons between different line outage strategies, the deterministic PSPS, in this section, optimizes commitment and operational decisions 
at the moment of peak expected demand, hour 16 in  Figure \ref{fig:Load_Profiles_IEEE14}, is considered.~\cite{Rhodes, Kody, Rhodes2, Kody2, Umunnakwe,Bayani} did not study generator commitment. Generator commitment statuses are analyzed in this paper because the selection of active generators highly influences total production costs. 
\begin{table}[ht]
\caption{Optimization approaches}
\begin{tabular}{ m{2em} m{7em} m{12.5em} } 
\hline
Name & Objective function & Risk constraint \\
\hline NMKS & ${f_{\text{uc}}} + f_{\text{VoLL}} + f_{\text{oc}}$ & $\sum_{\left({i,j} \right) \in \mathcal{L}} z_{ij,t} \leq |\mathcal{L}|-|\mathcal{K}|$ \\ 
MNWF & ${f_{\text{uc}}} + f_{\text{VoLL}} + f_{\text{oc}}$ &  Heuristic: activate k lines w/ k lowest $R_{ij,t}$ values\\
WFNC & $f_{\text{VoLL}}$  & $\sum_{\left({i,j} \right)} z_{ij,t}R_{ij,t} \leq R_{\text{tol}}$ \\ 
WFPI & ${f_{\text{uc}}} + f_{\text{VoLL}} + f_{\text{oc}}$ & $\sum_{\left({i,j} \right)} z_{ij,t}R_{ij,t} \leq R_{\text{tol}}$ \\  
WFSL & ${f_{\text{uc}}} + f_{\text{VoLL}} + f_{\text{oc}}$ & \makecell{$\sum_{\left({i,j} \right)} z_{ij,t}R_{ij,t} + R_{\text{slack}} = R_{\text{tol}}$, \\ $0 \leq R_{\text{slack}} \leq \overline{R}_{\text{slack}}$}\\ 
 \hline
\end{tabular}
\label{table:Benchmark}
\end{table}
%Approach $N$-$k$ (NMKS) simulates an approach to ensuring reliable operation after k number of lines of the grid are de-energized; however, NMKS is unaware of the line wildfire risks. 
In $N$-$k$ (NMKS), $|\mathcal{K}|$ lines are de-energized so that the total cost is minimized. Its drawback is that its optimization model is not aware of wildfire risk. The Minimum WFPI heuristic (MNWF) is the transmission heuristic approach described in \cite{Rhodes} and is commonly used in practice by system operators. Weighted WFPI with No Commitment or Operating Costs (WFNC) is a reformulation of the approach proposed in~\cite{Rhodes, Kody, Rhodes2, Kody2} in which total load served and wildfire risk are considered. The difference between WFNC and~\cite{Rhodes, Kody, Rhodes2, Kody2} is that allowable wildfire risk ($R_{\text{tol}}$) is treated as a constraint rather than an objective so as to keep all terms in the objective function in terms of financial costs. The weighted-WFPI approach (WFPI) has the same risk constraint as WFNC except that it considers commitment and operating costs in addition to VoLL. In Weighted-WFPI approach with Slack (WFSL), the WFPI approach is modified to include constraint~\eqref{eq:slack_max} that ensures the transmission line topology also optimizes wildfire risk rather than just selecting a feasible transmission line topology, which is a drawback of the WFPI approach. 

\begin{figure}[!ht]
\centering
\includegraphics[width=\columnwidth]{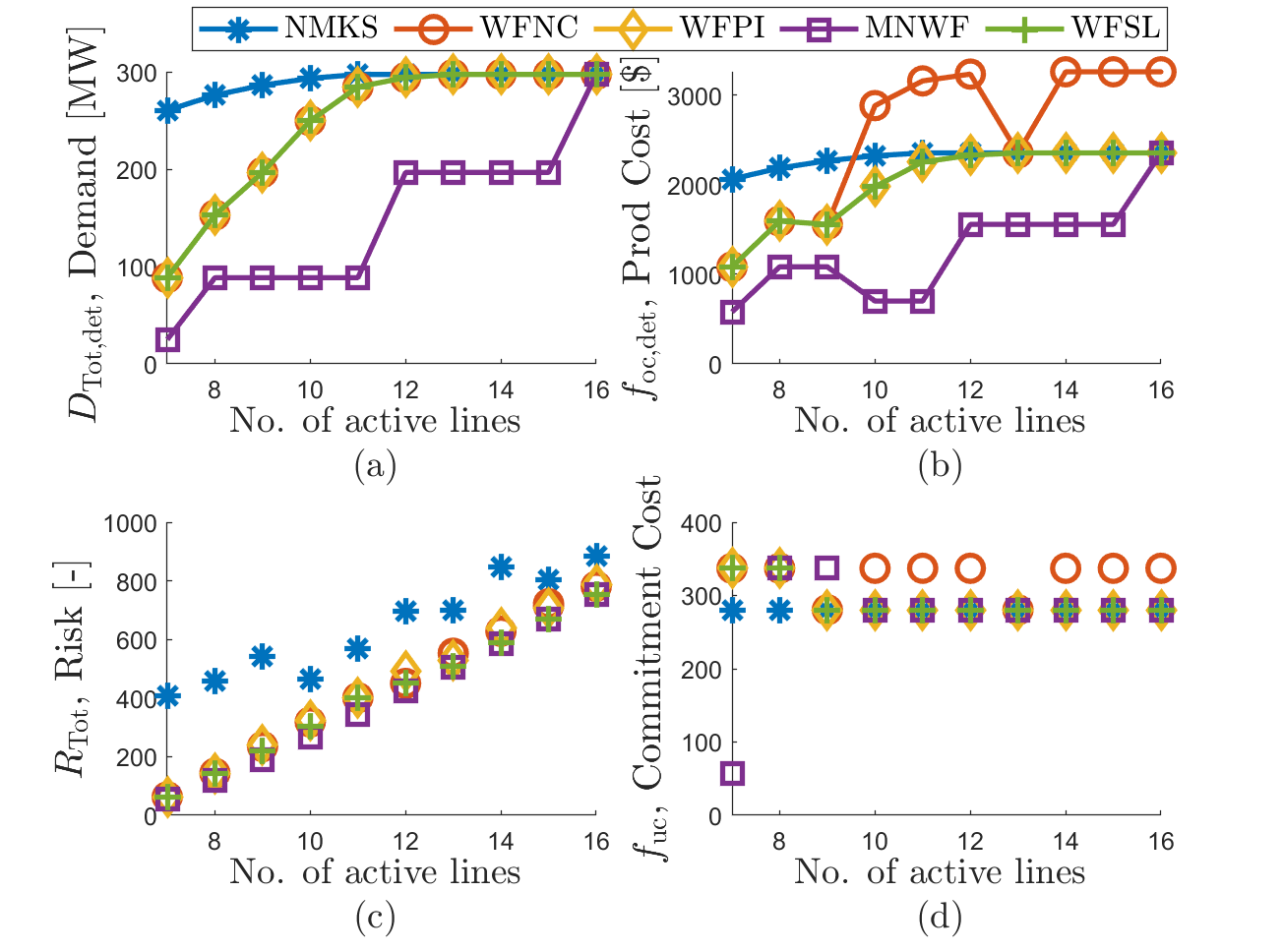}
\caption{Effect of the number of active lines on a) demand, b) production costs, c) wildfire risk, and d) commitment costs for the five benchmark optimization approaches described in Table~\ref{table:Benchmark} at the moment of peak demand (hour 16 of the day). %(b) shows that with a few exceptions, production costs tend to increase with the number of active lines; production costs increase for strategy WFNC which is unaware of production costs. (c) depicts a piece-wise linear increase in the cumulative wildfire line risk with increases in the number of active lines. %A highlight box shows that systems unaware of system fire risk could incur less fire risk than methods upper bounding risk with a fire risk tolerance. 
%(d) highlights that differences in objective functions and wildfire risk constraints can impose different restrictions on which generators can be started. Results from active line settings 0-6 are not shown because demand, production costs, and commitment costs for WFNC-WFSL are all equal to those generated by MNWF for 7 active lines. For more than 16 active lines, demand, production costs, and commitment costs are the same for all scenarios.
}
\label{fig:IEEE14DEPCLR}
%data: 
%Results.WFPIWC=load('worskpace_econWFPIWC101122IEEE14Bcases 16_Feb_2024_11_23_37.mat');
%Results.WFPINPC=load('worskpace_econWFPINPC101122IEEE14Bcases 16_Feb_2024_11_35_24.mat');
%Results.WFPISLK=load('worskpace_econWFPISLK101122IEEE14Bcases 16_Feb_2024_11_22_20.mat');
%Results.NMKS=load('worskpace_econNMKS101122IEEE14Bcases 16_Feb_2024_10_57_02.mat');
%Results.WFPI=load('worskpace_econWFPI101122IEEE14Bcases 16_Feb_2024_11_24_54.mat');
%script:PlotsCompareIEEE14T1v3.m
\end{figure}

Simulation results for all approaches are shown in Figure~\ref{fig:IEEE14DEPCLR}. For NMKS, $|\mathcal{K}|$ is swept from 0 to 20 active lines stepping one active line. For other approaches, $R_{\text{tol}}$ is swept from 0 to the sum of all 20 WFPI line values: 1,089. $R_{\text{tol}}$ is the set of minimum WFPI for a given number of active lines. Since 6 lines have zero WFPI, all approaches except NMKS reach the same solution for active line settings of 6 lines or less; that solution is to produce 26~MW at bus 2 using Gen 2.  

Due to the flexibility present in the IEEE 14-bus system, only 11 active lines are needed for NMKS to satisfy all of the demand and achieve the minimum cost of serving the maximum amount of demand (Figure~\ref{fig:IEEE14DEPCLR}(a)). However, NMKS achieves this solution at the cost of higher wildfire risk than other approaches. 13 active lines are needed for approaches WFNC, WFPI, and WFSL to serve all demand. With less than 13 active lines, total demand increases as the number of active lines increases. 
% $\textbf{General Trends in Production Costs:}$

The network structure greatly affects the production costs. While MNWF achieves the lowest production cost at each active line setting, it does this at the cost of producing the most load-shed compared to other approaches. And it is not until 16 lines are active that all of the demand is served. Additionally, MNWF cannot dispatch Gen 1 until at least 8 lines are active. This is because Gen 1 is located at bus 1 and the lines connected to bus 1 (Line 1-5 and Line 1-2) are the lines with the 8th and 10th smallest WFPI values in the network as shown in Fig.~\ref{fig:IEEE14colored}. MNWF needs at least 11 lines to be active before the heuristic decides to exclusively use the least costly generator (Fig.~\ref{fig:IEEE14DEPCLR}(d)). 

In WFNC, WFPI, and WFSL both generators need to be committed when there are 7 and 8 active lines because the resulting network is divided into two disjoint networks. 
% (see Figure~\ref{fig:IEEE14B8LBetaDETT24}). 
% One set of demands is served exclusively by Gen 1 and the other by Gen 2. 
Given the relative sizing and location of the less costly generator Gen 1, this generator is committed for active line settings 8-20 for all approaches. When more than 9 lines are active, only the less costly generator, Gen 1, needs to be dispatched for NMKS, WFPI, and WFSL as shown in Figure \ref{fig:IEEE14DEPCLR}(d); the commitment cost for those cases, \$280, is equal Gen 1's start-up cost.  

% \begin{figure}
% \centering
% \includegraphics[width=\columnwidth]{FiguresFolder/SamplePlots/IEEE14B7LBetaDETT24.png}
% \caption{Optimized deterministic 8-line shut-off strategy and average load delivery for the IEEE 14-Bus system to minimize commitment and production costs at the moment of peak demand (hour 16). The size and color of buses indicate the fraction of the load being served. Grey dashed lines indicate line de-energizations and black buses indicate no demand service.} 
% \label{fig:IEEE14B7LBetaDETT24}
% %data: worskpace_econWFPI101122IEEE14Bcases 02_Feb_2024_17_23_28
% %script:
% \end{figure}

% $\textbf{Why include production costs?}$

All Approaches achieve monotonic increases in demand with increases in active lines. NMKS has a monotonic increase in costs with an increase in active lines. WFPI and WFSL have similar trends in production costs versus active lines: increases in production costs from 6-8 active lines, a drop in costs with 9 active lines, then followed by a monotonic increase in production costs for 10 or more active lines (Fig.~\ref{fig:IEEE14DEPCLR}(b)). A drawback of not including production costs in WFNC is that there can be solutions that are more expensive yet meet the same demand with the same number of active lines. For example, in Fig.~\ref{fig:IEEE14DEPCLR}(b) the production costs of WFNC relative to NMKS, WFPI, and WFSL increase for active line setting of 10-12, 14-16 lines. In WFNC, the dispatch of the more costly second generator leads to production cost increases between roughly 39-46\% (or \$905 in case of 10-12, 14-16 active lines).
% $\textbf{Why include slack variable to tighten wildfire tolerance?}$

By constraining the number of active lines to the lowest amount of wildfire risk per line, MNWF displays the lowest possible wildfire risk; however, it reduces the load served for less than 16 active lines. Since wildfire risk is not included in the objective function, NMKS, WFNC, and WFPI are not guaranteed to generate active line strategies with the lowest wildfire risk for the same economic cost. There may be multiple line de-energization strategies that generate the same economic cost and are within the wildfire risk tolerance. For instance, given an $R_{\text{tol}}$ of 266.2, WFNC, WFPI, and WFSL serve 196.6 MWh at a cost of \$1,557.4 with 9 lines; 8 of those are in common (totaling a WFPI of 143.5). WFNC activates line 2-4 (77.1 WFPI), WFPI 2-5 (83.2 WFPI), and WFSL 1-5 (63.1 WFPI). NMKS improves load delivery by 45\% (or 286.3 MWh). However, it only activates 2 of the 6 zero WFPI lines (9-10 and 13-14) and increases wildfire risk by 145\% (541 WFPI). %Furthermore, as can be seen in Figure \ref{fig:IEEE14DEPCLR}(c), it is possible for the $N-k$ approach, which does not consider wildfire risk at all, to generate a solution with lesser wildfire risk than an approach that uses a wildfire risk weighted constraint. (see the case when there are 12 actives lines in Figure \ref{fig:IEEE14DEPCLR}(c)).
Approach WFSL generates lower wildfire risk as compared to NMKS. And WFSL generates lower wildfire risk than WFNC and WFPI despite serving the same amount of demand. This is accomplished by the slack variable that minimizes the difference from the minimum wildfire risk ($R_{\text{tol}}$) for a given number of active lines as described in Section~\ref{subsection: Constraint for Enacting No Less Than k Damaged Lines}. Another drawback to WFNC and WFPI is that finer sweeps of $R_{\text{tol}}$ may be needed to understand the relationship between $R_{\text{tol}}$ and the optimized number of active lines (a discrete variable). For instance, if the operator chooses an $R_{\text{tol}}$ of 56.89,  MNWF activates the six zero risk lines and line 6-13 (the line with the lowest non-zero risk of a line).
% As seen in Figure \ref{fig:IEEE14B7LBetaDETT24}, 
The addition of 6-13 does not improve the load delivery when both generators are isolated from the rest of the network. As a result of this zero benefit to the load delivery, WFNC and WFPI choose to only activate the six lines with zero WFPI until the WFPI tolerance is at least 63.10. WFSL increases total system line risk to WFPI 63.10 %) violates the $R_{\text{tol}}$ limit by 6.21 WFPI
to add line 1-5 instead of 6-13. While the system wildfire risk is increased by 10.9\%, load delivery is increased by 255\% (from 24.9~MW to 88.49~MW), which justifies Gen 1 commitment. 
%  Repeat for IEEE 14 Bus System and RTS GMLC
%  $t=1$ Compare demand, production cost, and WF Risk for Minimum WF Risk, WF Risk, WF Risk w/ slack and $N-k$, No prod cost w/o considering unit commitments

%  $t=1$ Stochastic Scenarios w/o considering unit commitment
% Considering prod cost could lead to lower production costs than not considering prod costs in objective
% At best not considering production costs will equal the optimal when considering production costs
% Considering wildfire risk could lead to lower cumulative wildfire risk than a strategy (such as $N-k$) unaware of WF risk. The solution when considering WF risk could be worse than without considering it in terms 
% Considering uncertainty in demand (and renewable resource availability) leads to commitments that can decrease expected production costs. See Figure \ref{fig:IEEE14DEPCLR}.

% \begin{figure}[h!]
% \centering
% \subfloat[]{\includegraphics[width=4.4cm]{FiguresFolder/SamplePlots/IEEE14DemandT1.png}}\hfil   
% \subfloat[]{\includegraphics[width=4.4cm]{FiguresFolder/SamplePlots/IEEE14ProdCostT1.png}}\hfil
% \subfloat[]{\includegraphics[width=4.4cm]{FiguresFolder/SamplePlots/IEEE14WFRiskT1.png}}
% \subfloat[]{\includegraphics[width=4.4cm]{FiguresFolder/SamplePlots/IEEE14CommitT1.png}}
% \caption{Plots of the optimized demand served (a), production costs (b), cumulative wildfire line risk (c), and generator commitments versus number of active lines. Plots (a)-(d) show results for different transmission line constraint options optimized at the moment of peak demand.} 
% \label{fig:IEEE14DEPCLR}
% \end{figure}
Overall Figure~\ref{fig:IEEE14DEPCLR} indicates that WFSL is the best approach of the five for deterministic PSPS. WFSL minimizes commitment, production, and VoLL costs and tightens the bounds of wildfire risk tolerance. The presence of the slack variable in WFSL allows operators to still use simple heuristics to step through wildfire risk tolerance levels. WFSL does not require a finer sweep of risk tolerances (as needed in WFNC or WFPI) to outperform an $N$-$k$ approach strategy in wildfire mitigation. 

\vspace{-1em}
\subsection{Stochastic PSPS (SPSPS) results for IEEE 14-Bus}
\label{subsection:Results for the SPSPS}

In the simulations, $\epsilon$ is fixed at the 95\% confidence level and the VoLL is set to 1,000 \$/MWh \cite{Trakas2,Mohagheghi,Farzin}. Since the value of losing 1~MWh is at least one order magnitude higher than the cost to produce 1 MWh, the VoLL is typically the largest contributor to the total economic costs. For simulations, the step size of $\beta$ is taken as 0.1; all samples are summarized by $\beta \in \{ 0,0.1,\dots,1\}$.

\subsubsection{Sweep of risk aversion levels for IEEE 14-Bus system}
% optimized at moment of peak demand}
% to high economic costs} 
\label{subsubection:Sweep of Risk Aversion Levels} 

To highlight the cost differences between the risk-neutral and risk-averse operating strategies, the case when the operator selects 12 active lines is considered at moment of peak demand.
% By definition, purely risk-neutral optimal decisions ($\beta=0$) only minimize expected costs while purely risk-averse optimal decisions ($\beta=1$) only minimize CVaR costs. For all active line settings, all stochastic first-stage generator commitment and line de-energization decisions generated with $\beta$ between 0 and 1 were either equal to those with $\beta=0$ or with $\beta=1$. For the sake of generalizing the results of the two-stage stochastic optimization, we classify decisions stemming from the two-stage stochastic solution into two categories: risk-neutral and risk-averse. The risk-neutral strategies contain a $\beta$ value that generates the same first-stage decisions as the $\beta=0$ case (which is risk-neutral by definition) and risk-averse strategies contain a $\beta$ value that generates the same first-stage decisions as the $\beta=1$ case. 
The risk-neutral strategies correspond to $\beta \in [0,0.1]$ and $\beta \in [0.2,1]$ leads to risk-averse strategies. Optimal expected and CVaR total costs, production costs, total demand served, and generator commitments are given in Table~\ref{table:12ActiveLines}. It should be emphasized that only the first-stage generator commitment and line de-energization decisions are used in the second stage while the generator powers and load shed are discarded and re-optimized for a particular realization of demand in the second stage.   

\begin{figure}[!ht]
\centering
% \subfloat{\includegraphics[width=4.426cm]{FiguresFolder/SamplePlots/IEEE14B12T1BarPgCompare.png}}
% \hfil
% \subfloat{\includegraphics[width=4.426cm]{FiguresFolder/SamplePlots/IEEE14B12T1BarLoadCompare.png}}
% \hfil
% \subfloat{\includegraphics[width=4.426cm]{FiguresFolder/SamplePlots/IEEE14B12T1BarCgCompare.png}}
% \hfil
% \subfloat{\includegraphics[width=4.426cm]{FiguresFolder/SamplePlots/IEEE14B12T1BarfVoLLCompare.png}}
\includegraphics[width=\columnwidth]{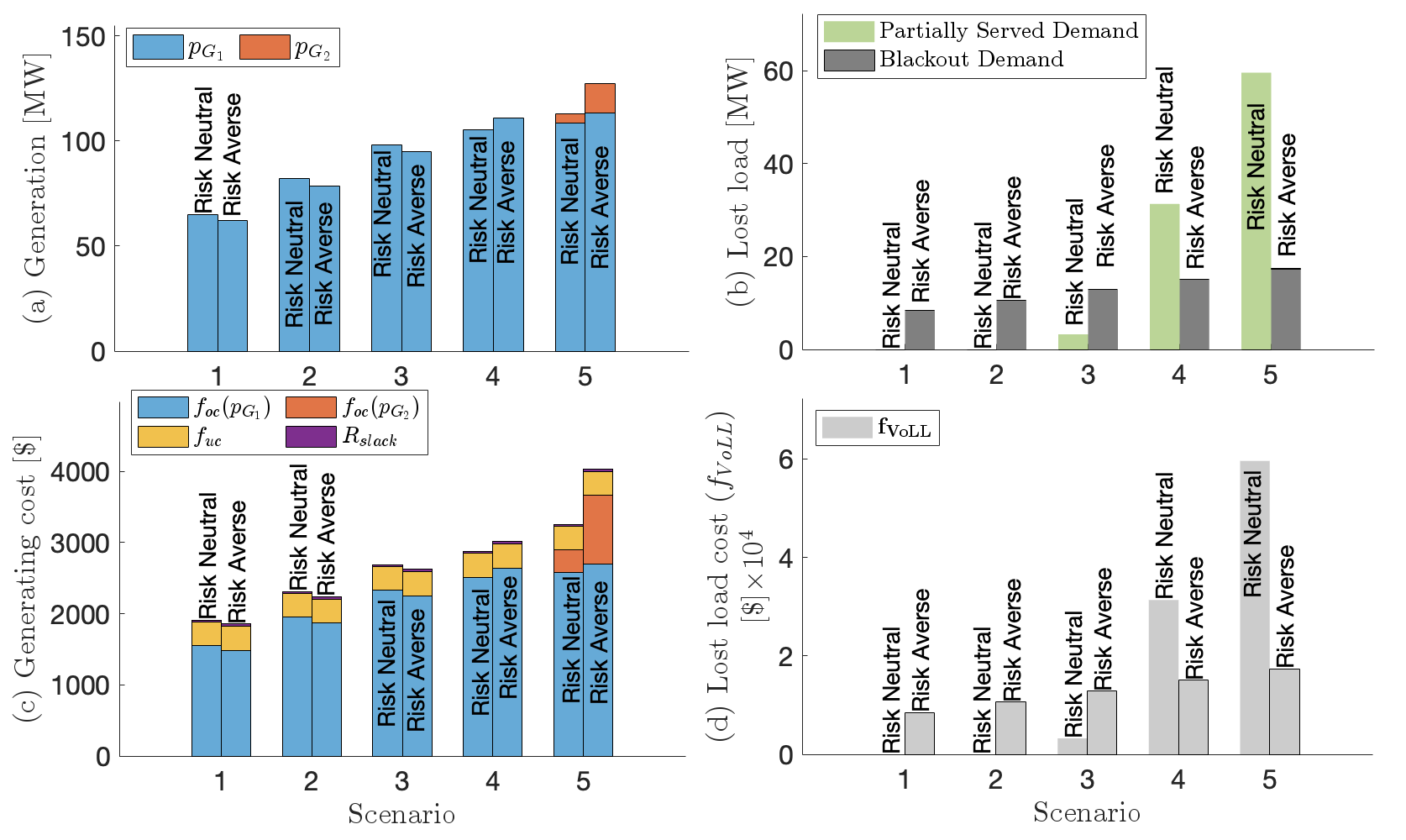}
\caption{SPSPS (a) generation, (b) load shedding, (c) total costs (excluding VoLL), and (d) VoLL costs for the IEEE 14-bus system optimized at the moment of peak demand (hour 16). Left bars are risk-neutral results for $\beta=0$ and right bars are risk-averse results for $\beta=0.8$. 
% (a) depicts more generation by the risk-neutral approach in the three lower demand scenarios and less generation in the two higher scenarios than the risk-averse approach. 
% Both strategies activate the second generator only in the highest demand scenario. 
% (b) depicts more lost load due to partially served demand in risk-neutral cases and more lost load due to blackouts in risk-averse cases. 
% (c) shows total generating cost. (d) depicts how the lost load in (b) is scaled by the VoLL. 
Subplots (a)-(d) together show that for the risk-averse strategy, higher production costs in scenarios 4 and 5 are overcome by savings in VoLL.
} 
\label{fig:BarGenCostCompare12L}
%data: worskpace_econWFPI101122IEEE14Bcases 02_Feb_2024_17_23_28
% Consult variable OptTimeTable
% Last column comes from PI_TOTAL1 and PI_TOTAL_EEV (last entry last column)
%script:fig6BarGenCostCompare12L.m
\end{figure}

Figure~\ref{fig:BarGenCostCompare12L} summarizes differences in the generation, lost load, generating costs, and VoLL costs that arise between risk-neutral ($\beta \in [0,0.1]$) and risk-averse ($\beta\in [0.2,1]$) decisions. Note that the solutions for $\beta=0.8$ and $\beta=0$ are the ones plotted but any $\beta$ within the respective risk-averse or risk-neutral ranges results in the same first-stage generator commitments, line de-energizations, and second-stage decisions. 

% \begin{figure}[h!]
% \centering
% \subfloat[]{\includegraphics[width=\columnwidth]{FiguresFolder/SamplePlots/IEEE14B12T1BarGenCompare.png}}\hfil
% \subfloat[]{\includegraphics[width=\columnwidth]{FiguresFolder/SamplePlots/IEEE14B12T1BarCostCompare.png}}
% \caption{Bar Charts of the (a) generation dispatch, lost load, (b) total generating costs, and VoLL costs for the risk-neutral $\beta=0$ and risk-averse $\beta=0.8$ cases for each of the 5 scenarios at the moment of peak demand} 
% \label{fig:BarGenCostCompare}
% %DATA: worskpace_econWFPI101122IEEE14Bcases 31_Jan_2024_10_59_21
% \end{figure}

\begin{table*}[!ht]
\begin{center}
\caption{SPSPS commitment and operational decisions and costs for the IEEE 14-bus system optimized at the moment of peak demand (hour 16). 
% Cumulative transmission line WF risk (WFPI), 1st stage expected demand, 1st stage $\text{CVaR}_{0.95}$ demand, 1st stage expected production cost,  1st stage $\text{CVaR}_{0.95}$ production cost, \# of generator commitments, total 1st stage expected cost, 1st stage $\text{CVaR}_{0.95}$ costs, and the 2nd stage cost if the highest demand scenario is realized for 12 actives lines. 
$\beta$ is stepped every 0.1 from $\beta=0$ (risk neutral) to $\beta=1$ (most risk-averse). CVaR is calculated using equation $(7)$ from \cite{Sarykalin}.
}
%\newcolumntype{A}{>{$\displaystyle}>{\raggedleft\arraybackslash}m{1.0cm}<{$}}
\begin{tabular}{l  
>{\raggedleft\arraybackslash}m{1.0cm}
>{\raggedleft\arraybackslash}m{1.5cm} 
>{\raggedleft\arraybackslash}m{1.5cm}  
>{\raggedleft\arraybackslash}m{1.2cm} 
>{\raggedleft\arraybackslash}m{1.5cm}  
>{\raggedleft\arraybackslash}m{1.8cm}  
>{\raggedleft\arraybackslash}m{1.0cm} 
>{\raggedleft\arraybackslash}m{1.1cm} 
>{\raggedleft\arraybackslash}m{1.4cm}
}
\hline  Risk Aversion & Line Risk &  Exp. Demand [MWh]& $\text{CVaR}_{0.95}$ Demand [MWh] & Exp. Prod. Cost [\$] & $\text{CVaR}_{0.95}$ Prod. Cost [\$] & Gen Commits & Exp. Cost [\$] & $\text{CVaR}_{0.95}$ Cost [\$] & Highest Demand Cost [\$]\\
\hline 
$\beta=1$ & 464.3 & 281.8& 355.2& 2,243.7 & 3,104.0 &2 & 18,106 & 19,551 & 21,408 \\
$\beta \in [0.2,0.9]$ & 464.3 & 284.4 & 355.2 & 2,267.5 & 3,104.0 & 2 & 15,463 & 19,551 & 21,408
\\  
$\beta \in [0,0.1]$ & 450.0 & 289.4& 
 327.1 & 2,297.4 & 2,684.3 & 2 & 10,472 & 47,213 & 62,824 \\
\text{Deter.} & 450.0 & 294.0 & - & 2,329.0 & - & 1& 5,863 & - & 65,931 \\
\hline
\end{tabular}
\label{table:12ActiveLines}
\end{center}
%data: worskpace_econWFPI101122IEEE14Bcases 02_Feb_2024_17_23_28
%script:
%tips on table construction https://www.tug.org/pracjourn/mori/tables.pdf
\end{table*}

In the case of 12 active lines, the risk-neutral decision maker commits the less costly Gen 1 until it becomes cost-effective to commit Gen 2 in the highest-demand scenario. The peak demand is equal to the maximum combined capacity of the two generators; however, transmission line power flow limits lead to unserved energy in the three higher demand scenarios. In the highest demand scenario, lines 2-5 and 2-3 reach their power flow limits. The risk-neutral approach not only uses less of Gen 1's capacity but also less combined generator capacity in the highest demand scenario than the risk-averse approach (see Fig.~\ref{fig:BarGenCostCompare12L}(a) for comparison of generation and Table \ref{table:DiffLineShutoffAveDemand} for differences in transmission line de-energization). VoLL costs are purely a function of the risk-neutral decisions maxing out power flow line limits during the three highest demand scenarios (see Figs. \ref{fig:BarGenCostCompare12L}(b) and \ref{fig:BarGenCostCompare12L}(d)). 

The risk-averse objective of minimizing CVaR total economic costs is accomplished by minimizing a weighted average of costs among the top 5\% of scenario costs. Given the distribution of costs in Fig.~\ref{fig:Load_Profiles_IEEE14}, the weighted average is 54.4\% of the total cost of the fourth scenario and 45.6\% of the total cost of the fifth scenario. As seen in Figure~\ref{fig:BarGenCostCompare12L}(c)), this weighted average of costs in the fourth and fifth scenarios is less for the risk-averse decisions than it is for the risk-neutral decisions. In the 12 active lines setting, the risk-averse approach is accomplished by altering the network configuration (see Table~\ref{table:DiffLineShutoffAveDemand}) to achieve lower conditional expected total costs among the two highest demand scenarios. Power flows are not exceeded in any of the lines in the four lower demand scenarios, but the alteration of the network blackouts bus 6 for all scenarios. There is a small amount of 1.16~MW loss in the load at bus 3, $p_{D_3}$, (0.8\% of 145~MW load) in the highest demand scenario for the risk-averse decisions. Therefore, the VoLL in each scenario for the risk-averse decision maker is mostly due to the blackout at bus 6. This effect is also observed in the Section~\ref{subsubsection:24hr Representative Day Commitment and Analysis} which considers the full 24-hour horizon. Specifically, the risk-averse operators reduce the expected costs of operation among the high-demand scenarios at the expense of losing more demand buses to blackouts.   

Table~\ref{table:12ActiveLines} shows that the risk-averse decision to blackout bus 6 leads to less expected production costs; however, the blackout of bus 6 in each load scenario increases expected total costs through higher expected VoLL costs. In the 12 active line setting, the risk-averse decisions lead to more $\text{CVaR}_{0.95}$ production costs than risk-neutral decisions. However, the increase in CVaR production costs is due to serving a greater portion of the load; less lost load equates to less lost load costs and less total cost in the two high-demand scenarios.

\begin{figure}[!ht]
\centering
%\subfloat[]
\includegraphics[width=\columnwidth]{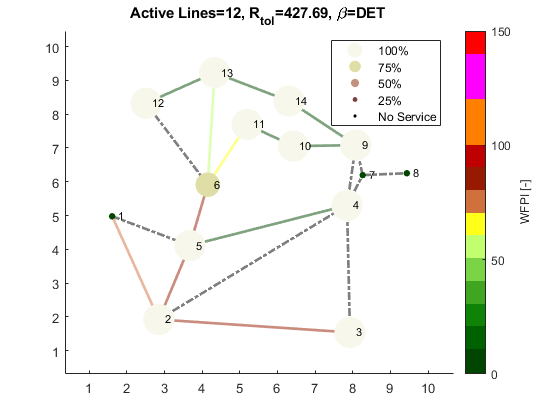}
%\hfil 
% \subfloat[][]{
% \begin{tabular}{l m{1.5cm} m{1.6cm} m{1.6cm}}
% \hline 
%  $z^{\beta}_{i,j}=1$  &$>z^{\text{DET}}_{ij}$ & $<z^{\text{DET}}_{ij}$ \\
% \hline 
% $\beta=1$       & $z_{1,5},z_{2,4}$ & $z_{2,5},z_{5,6},z_{6,11},z_{6,13}$ \\
% $\beta=0.2-0.9$ & $z_{1,5},z_{2,4}$ & $z_{2,5},z_{5,6},z_{6,11},z_{6,13}$ \\  
% $\beta=0-0.1$   & N/A               & N/A                                 \\
% \hline
% $\mathbb{E}_{\omega}[p_{d,t}(\xi)]$ &$>p^{\text{DET}}_{d,t}(\bar{\xi})$& $<p^{\text{DET}}_{d,t}(\bar{\xi})$ & $=0$\\
% \hline
% $\beta=1$       & $p_{D_4,t}$                    & $p_{D_5,t}$                  & $p_{D_6,t}$ \\
% $\beta=0.2-0.9$ & $p_{D_4,t}$                    & N/A                          & $p_{D_6,t}$ \\  
% $\beta=0-0.1$   & $p_{D_4,t}$                    & $p_{D_{10},t},p_{D_{14},t}$      & N/A \\
% \hline
% \end{tabular}
% }
\caption{Optimized deterministic 12-line shut-off strategies and load delivery to minimize commitment, production costs, and VoLL at the moment of peak demand. The size and color of buses indicate the fraction of the load being served. Black dashed lines indicate line de-energizations and black dot buses indicate no demand served.}
\label{fig:IEEE14B12LBeta}
%data: worskpace_econWFPI101122IEEE14Bcases 02_Feb_2024_17_23_28
%script: fig4_7_IEEE14B12LBeta.m
\end{figure}

The instance when the number of active lines equals 12 is a notable case because the most risk-averse ($\beta=1$), risk-averse ($\beta \in [0.2,0.9]$), risk-neutral ($\beta\in[0,0.1]$), and deterministic approaches produce different total expected costs based on the level of risk aversion. The network for the deterministic case is depicted in Figure~\ref{fig:IEEE14B12LBeta} to give the reader a baseline for comparison to strategies developed by the two-stage stochastic solution at different risk averseness levels.

Figure~\ref{fig:IEEE14B12LBeta} and Table~\ref{table:DiffLineShutoffAveDemand} summarize the differences in transmission line de-energizations and expected demand served at various buses. At the bus level, compared to the deterministic case, there is a reduction in the expected demand at buses 4, 9, 10, and 13 for $\beta$ from 0 to 0.1 (more risk neutral). %10 and 14. 
This reduction in expected demand is due to power flow limits on line 5-6 restricting service to buses downstream of bus 6 %buses 10 and 14. 
in the higher demand scenarios of the risk-neutral optimization.  

Then going from $\beta \in [0, 0.1]$ to $\beta \in [0.2, 1]$ (increasing in risk-averse level), there is a restoration of the demand at buses 4, 9, 10, and the lines 1-5, 2-4, 4-7, and 7-9. But there is a blackout at bus 6 and shut-offs of lines 2-5, 5-6, 6-11, and 6-13. As mentioned previously, this new configuration allows more of the capacity in Gen 1 and Gen 2 to be used. Finally, in the case when $\beta=1$ (most risk-averse case), there is an additional reduction in the expected demand at bus 4 since only the $\text{CVaR}_{0.95}$ term is considered in the optimization problem. The $\text{CVaR}_{0.95}$ term optimizes costs in the two most costly demand scenarios (e.g. the fourth and fifth demand scenarios). Additional load loss in the three less costly lower demand scenarios occurs because those cost scenarios are not optimized. %compared to demand scenarios 4 and 5.

\begin{table}[!ht]
\begin{center}
\caption{Differences between stochastic line de-energizations ($z^{\beta}_{ij}$) and the deterministic line de-energization ($z^{\text{DET}}_{ij}$). Differences between expected load served, $\mathbb{E}[p^{\beta}_{d,\bm{\omega}}]$, and deterministic load served, $p^{\text{DET}}_{d}(\mathbb{E}[\bm{\xi}])$. Ranges of risk averseness are $\beta \in [0,0.1]$ and $\beta \in [0.2,1]$. The increase in risk averseness levels leads to blackouts at Bus 6 and decreases in expected demand}
\begin{tabular}{c p{1cm} p{0.9cm} p{0.9cm} p{0.9cm} p{0.9cm} p{0.9cm}}
\hline 
${\beta}$  &  \multicolumn{3}{p{2.7cm}}{$z^{\beta}_{ij}=1 $, $z^{\text{DET}}_{ij}=0$} & \multicolumn{3}{p{2.7cm}}{$z^{\beta}_{ij}=0 $, $z^{\text{DET}}_{ij}=1$} \\
% $z^{\beta}_{ij}=1$  &$>z^{\text{DET}}_{ij}$ & $<z^{\text{DET}}_{ij}$ \\
\hline 
%$\beta=1$       & $z_{\{15,24,47,79\}}$ & $z_{\{25,56,611,613\}}$ \\
$0.2-1$ & \multicolumn{3}{c}{$z_{\{15,24,47,79\}}$} & \multicolumn{3}{c}{$z_{\{25,56,611,613\}}$} \\  
$0-0.1$   & \multicolumn{3}{c}{-}               & \multicolumn{3}{c}{-}                                 \\
\hline
$\beta$ & \multicolumn{2}{p{1.8cm}}{$\mathbb{E}[p^{\beta}_{d,\bm{\omega}}]$$>p^{\text{DET}}_{d}(\mathbb{E}[\bm{\xi}])$}& \multicolumn{2}{ p{1.8cm} }{$\mathbb{E}[p^{\beta}_{d,\bm{\omega}}]$$<p^{\text{DET}}_{d}(\mathbb{E}[\bm{\xi}])$} & \multicolumn{2}{p{1.8cm} }{Blackout demand}\\
\hline
$1$       & \multicolumn{2}{c}{-}                   & \multicolumn{2}{c}{$p_{D_4}$}                  & \multicolumn{2}{c}{$p_{D_6}$} \\
$0.2-0.9$ & \multicolumn{2}{c}{-}                    & \multicolumn{2}{c}{-}                          & \multicolumn{2}{c}{$p_{D_6}$} \\  
$0-0.1$   & \multicolumn{2}{c}{$p_{D_6}$}                    & \multicolumn{2}{c}{$p_{D_{\{4,9,10,13\}}}$} 
 %$p_{D_{4},t},p_{D_{9},t},p_{D_{10},t},p_{D_{13},t}$%      
& \multicolumn{2}{c}{-} \\
\hline
\end{tabular}
\label{table:DiffLineShutoffAveDemand}
\end{center}
%data: Different data than worskpace_econWFPI101122IEEE14Bcases 02_Feb_2024_17_23_28
%script: Look at z_l_TOTAL and x_d_TOTAL10ave

\end{table}

% \begin{table}[h!]
% \begin{center}
% \caption{Increase in risk averseness to 0.95-quantile total costs leads to different line statuses and decreases in average 1st stage demand}
% \begin{tabular}{l c c m{0.6cm}}
% \hline 
% $z^{\beta}_{i,j}=1$  &$>z^{\text{DET}}_{ij}$ & $<z^{\text{DET}}_{ij}$ \\
% \hline 
% $\beta=1$       & $z_{15},z_{24},z_{47},z_{79}$ & $z_{25},z_{56},z_{611},z_{613}$ \\
% $\beta=0.2-0.9$ & $z_{15},z_{24},z_{47},z_{79}$ & $z_{25},z_{56},z_{611},z_{613}$ \\  
% $\beta=0-0.1$   & -               & -                                 \\
% \hline
% $\mathbb{E}_{\omega}[p_{d,t}(\xi)]$ &$>p^{\text{DET}}_{d,t}(\bar{\xi})$& $<p^{\text{DET}}_{d,t}(\bar{\xi})$ & $=0$\\
% \hline
% $\beta=1$       & $p_{D_4,t}$                    & $p_{D_5,t}$                  & $p_{D_6,t}$ \\
% $\beta=0.2-0.9$ & $p_{D_4,t}$                    & -                          & $p_{D_6,t}$ \\  
% $\beta=0-0.1$   & $p_{D_4,t}$                    & $p_{D_{10},t},p_{D_{14},t}$      & - \\
% \hline
% \end{tabular}
% \label{table:DiffLineShutoffAveDemand}
% \end{center}
% %data: Different data than worskpace_econWFPI101122IEEE14Bcases 09_Feb_2024_17_23_28
% %script: Look at z_l_TOTAL and x_d_TOTAL10ave

% \end{table}

Table~\ref{table:12ActiveLines} follows the anticipated trend that increasing $\beta$ increases mean total costs and decreases $\text{CVaR}_{0.95}$ total costs (explained in the last paragraph in Section \ref{subsection:Two Stage Stochastic Problem}.1). The expected costs for the risk-averse strategies ($\beta \in [0.2,0.9]$ and $\beta=1$) are nearly \$5,000 and \$7,700 higher than the risk-neutral approach. On the other hand, $\text{CVaR}_{0.95}$ costs for those more risk-averse approaches are roughly \$27,500 less than the risk-neutral approach. There is significantly more load shed in the higher demand scenarios than the lower demand scenarios (more than 5~MW) which results in higher $\text{CVaR}_{0.95}$ costs for the risk-neutral approaches. 

The deterministic strategy produces less expected cost than the risk-neutral strategy. However, if the realized demand is the highest demand scenario, only committing Gen 1 in the second stage (as done in the deterministic problem) instead of both generators equates to %roughly a \$75 difference in VoLL costs and 
nearly \$3,107 in total costs (last column of Table~\ref{table:12ActiveLines}) or %The deterministic commitment decisions equate to an increase in the total cost of \$1,087; that's 
an increase of \$15 in expected total economic costs. %after subtracting the \$56 spent to commit the second generator and additional production costs due to serving an additional 3 MW generator by gen 2 and 10 MW generated by the gen 2 that could have been generated by gen 1 in the highest demand scenario. 
The \$15 reduction in expected total costs shows that there is a financial value added in considering the two-stage stochastic unit commitments over the deterministic problem. Finally, we also carried out simulations for different number of scenarios ranging from 2 to 100 for deterministic and risk-neutral cases only. Increasing the number of scenarios improves the optimal cost but increases the solve time, particularly for cases with less than 11 active lines which is the minimum number of lines required to serve all of the demand. Selecting an appropriate number of scenarios is a trade-off between solve time and optimal cost and would depend upon the computation resources available to the operator.

\subsubsection{24hr representative day analysis}
\label{subsubsection:24hr Representative Day Commitment and Analysis}
The case when there are 11 active lines is analyzed for the full 24-hours for a representative day. Since the deterministic case only considers an expected demand with a maximum of 297~MW, which is less than the capacity of Gen 1, the second generator is not committed. Similar results were observed in Fig.~\ref{fig:IEEE14DEPCLR}(d). The results for the stochastic cases are shown in Figure~\ref{fig:IEEE1411LCommitLoadsto24h}(a)-(c) for the least demand scenario (scenario 1) and two highest demand scenarios (scenarios 4 and 5).
% Compared to the case with 12 active lines in the previous section in which only the highest demand was considered, there is no change in network configurations, commitments, and line de-energization decisions for the 24 hour period. 
% De-energization strategies are similar to risk-neutral and risk-averse strategies with 11 active lines optimized for 24h as they are for 12 active lines optimized for only the peak hour. 
% This is because the de-energization constraint \eqref{eq:Damaged} and WFPI values are constant throughout the day.
% , de-energization decisions tend to be constant throughout the day. 
% The difference between the deterministic and risk-neutral de-energizations and Figure \ref{fig:IEEE14B12LBeta} is that line 6-13 is not active. The risk-averse case chooses to blackout bus 6 and energize lines 4-7 and 4-9 instead of 5-6, 6-11, and 6-13. The risk-averse strategy drops the demand at bus 6 but, it adds a maximal power flow from bus 4 to bus 9 of 87 MW and splits the rest of the network into two branches( a branch from 9-11 and 9-12). Opening lines 5-6 and 6-11 adds a maximal flow capacity of 63 MW from bus 5 to bus 11 and some of that flow is absorbed by the demand at 6. With the risk-neutral and deterministic line de-energizations, the rest of the network is a single branch from bus 5 to bus 12. The buses at the ends of that branch, buses 12, 13, and 14, are typically under-served. In the high-demand scenario, at least one of those three buses is under-served from the 7th to 24th hour (see Figure \ref{fig:IEEE1411LCommitLoadsto24h}(d)).   
In Fig.~\ref{fig:IEEE1411LCommitLoadsto24h}(a)-(c), the second generator is committed during the peak hours of the day (14-17 hours) and the corresponding time window is highlighted in green. Fig.~\ref{fig:IEEE1411LCommitLoadsto24h}(a)-(c) shows that the risk-averse cases tend to serve more demand during peak periods of the day than during hours of less demand as compared to the risk-neutral approaches. As discussed in Section~\ref{subsection:Results for the SPSPS}.1, the definition of $\text{CVaR}_{0.95}$ and the allocation of the same VoLL to each bus regardless of the location and time of day contribute to large drops in the three less costly first-stage demand scenarios for $\beta=1$. The black dashed lines represent the demand corresponding to the five scenarios used in the optimization. Finally, for all the stochastic optimization strategies and the deterministic case, some portion of the expected total load is not served between hours 11 and 21. 
% Similar to what was seen in Table \ref{table:12ActiveLines}, the expected demand served for the stochastic problem (colored dashed lines) is below the expected demand from the deterministic case during the 11th to 20th hours due to reductions in the demand served in higher-demand scenarios. 
% The black dashed line indicates that between the 11th and 21st hours, some portion of the expected total load is not being served regardless of the optimization strategy. 
\subsubsection{Second stage results}
\label{subsubsection:second stage}
In the second stage, the line de-energization and generator commitment decisions from the first stage are used to solve a 24 hour receding horizon optimization problem first for scenario 4 (S4) and then for scenario 5 (S5) as described in Section~\ref{subsubsection:Second Stage Formulation}. %Scenario 5 has higher demand than scenario 4. 
The results are tabulated in Table~\ref{fig:SecondStage24hCost:CommitGen}. The risk-averse case has the lowest $\text{CVAR}_{0.95}$, but the demand served and total cost behavior varies as explained next. In scenario 4, the risk-averse case results in higher demand served ($5,710$~MWh) for lower cost (\$$318,620$) than the risk-neutral case ($5,704$~MWh, \$$324,110$). The opposite trend is observed for scenario 5 where the risk-averse case results in lower demand served ($6,533$~MWh) for higher cost (\$$572,330$) than the risk-neutral case ($6,357$~MWh, \$$568,400$). The deterministic case performed better than the risk-neutral approach in scenario 4 ($5,704$ MWh, \$$324,050$) since it does not activate Gen 2. However, in scenario 5, the deterministic performance ($6,347$ MWh, \$$577,320$) is more costly and serves less demand than the stochastic cases. 
% Despite the risk-averse case having the lowest $\text{CVaR}_{0.95}$ costs, if the demand realization lies in the highest demand scenario, the risk-neutral case generates the least costs, while a demand realization in the second highest demand scenario resulted in the risk-averse case generating the lowest costs. Also, the deterministic case serves the same amount of load as the risk-neutral case for less cost in the fourth scenario because it does not activate the second generator.

\begin{figure}[!ht]
\centering
% \subfloat{\includegraphics[width=4.426cm]{FiguresFolder/SamplePlots/IEEE1424hGenCommitL11mod2.png}} 
% %\hfil
% \subfloat{\includegraphics[width=4.426cm]{FiguresFolder/SamplePlots/IEEE1424hLoadTotalL11mod2.png}
% }  
% \hfil
% \subfloat{\includegraphics[width=4.426cm]{FiguresFolder/SamplePlots/IEEE1424hLoadTotalL11Highmod.png}
% }
% % \hfil
% % \subfloat[]{\includegraphics[width=4.3cm]{FiguresFolder/SamplePlots/IEEE1424hXdL11.png}
% % }
% %\hfil
% \subfloat{\includegraphics[width=4.426cm]{FiguresFolder/SamplePlots/IEEE14B12T24BarLoadCompareScen5.png}}
\includegraphics[width=\columnwidth]{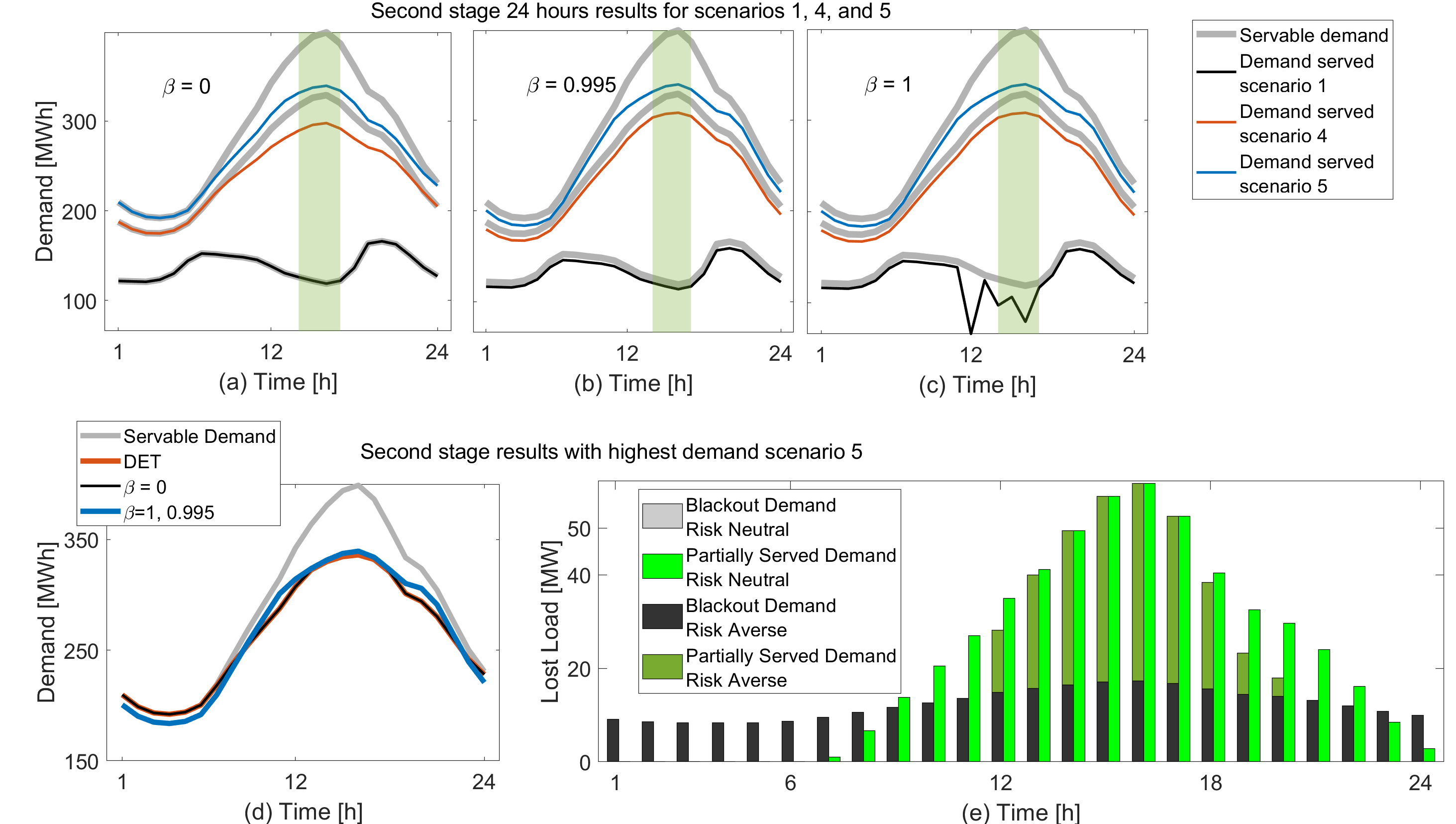}

\caption{All results are for 11 active lines at different risk averseness levels.
% (a) Optimized Gen 2 commitments; Gen 1 is always active. 
Network-wide load served for demand scenarios 1, 4, and 5 for $\beta = 0$ (a), $\beta = 0.995$ (b), $\beta = 1$ (c). Green highlights the time window shows where Gen 2 is active.
% (i.e. $\sum_{d \in 
% \mathcal{D}} x^{\beta}_{d,t,\omega}$). 
% Each point along a scenario load delivery timeseries is the $\sum_{d \in 
% \mathcal{D}} x^{\beta}_{d,t,\omega}$ for a particular timestep $t$.
(d) Second-stage network-wide load served when the realized demand is the highest demand (scenario 5). %(i.e. $\sum_{d \in \mathcal{D}} x^{\beta}_{d,t,\omega_{5}}$).
% (c) Note increases in the second-stage total load served for the high-demand scenario during hours 9-13 and 18-23 as risk averseness increases. Each point on a timeseries is the $\sum_{d \in \mathcal{D}} x^{\beta}_{d,t,\omega_{5}}$ for a particular timestep $t$. 
(e) Types of load shed for the risk-neutral and risk-averse strategies as blackout demand or partially served demand.
%(d) depicts a bar chart of the fraction of demand being served. Each cube is the $\sum_{t=1}^{24} x^{\beta}_{d,t,\omega}$ for a particular node $d$ and particular risk-averseness level $\beta$. Demands are sorted so that blue represents $p_{D_{2},t}$ and yellow $p_{D_{14},t}$  
} 
\label{fig:IEEE1411LCommitLoadsto24h}
%data: worskpace_econWFPI101122IEEE14Bcases 09_Feb_2024_20_21_12.mat
%script:fig9IEEE1411LCommitLoadsto24h.m
\end{figure}

The demand served and lost load for scenario 5 are plotted in Fig.~\ref{fig:IEEE1411LCommitLoadsto24h}(d) and (e) respectively. The line energization decisions in the risk-neutral and deterministic cases are the same, but differ from the risk-averse case. Specifically, in the risk-averse case, the lines connected to bus 6 are not energized which results in the demand at that node being blacked out for all hours of the day (Fig.~\ref{fig:IEEE1411LCommitLoadsto24h}(e)). The partially served demand at all buses (except bus 2, bus 3, and bus 6) is due to the power flow limit at line 2-5 which is reached from hours 12 to 20. Similarly, line 2-3 reaches its power flow limit at hour 16, affecting load delivery for bus 3. Over the entire time horizon, aside from the blackout at bus 6, buses 12 and 13 are the second and third most affected buses and lose nearly 16\% and 13\% of their total 24-hr load respectively. In the risk-neutral case, some portion of the demand is always served at all nodes. Buses 2, 4, 5, and 11 are at full service throughout the time horizon. Line 2-5 reaches its limit from hour 14 to 20 and line 2-3 in hour 16 resulting in unserved demand at bus 3. Buses 12 and 14 are the two most affected losing 34\% and 32\% of their total 24-hr load respectively. Since the deterministic case commits only Gen 1, the network cannot receive additional support from Gen 2 to serve buses 2 and 3 during hours 14 to 17 when line 2-5 reaches its maximum power flow capacity. Buses 2, 4, 6, and 10 had full service throughout the optimization horizon; however, buses 12 and 13 lost 70\% and 44\% of their total 24-hr load. 

\begin{table}[!ht]
\begin{center}
\caption{Demand served and cost of the second stage optimization with $\text{VoLL}$ equal to $1,000$~\$/MWh for Scenarios 4 and 5.}
\begin{tabular}{p{1.3cm} p{1cm} p{1cm} p{1cm} p{1cm} p{1cm}}
\hline Risk averse ($\beta$) & S4 Dem. [MWh] & S4 Cost [\$]& S5 Dem. [MWh]& S5 Cost [\$] & $\text{CVaR}_{0.95}$ Cost [\$] \\
\hline 
$0.995-1$ & 5,709.8&  318,620
& 6,352.7 &572,330 & 434,060\\
$0-0.995$ & 5,704.3 & 324,110
 & 6,356.6 & 568,400 & 435,260\\  
\text{Deter.} & 5,704.3 & 324,050
 & 6,347.0 &577,320 & 439,290\\
\hline
\end{tabular}
\label{fig:SecondStage24hCost:CommitGen}
\end{center}
%data: worskpace_econWFPI101122IEEE14Bcases 09_Feb_2024_20_21_12.mat
%script: PI_TOTAL1 for Cost rows 1 and 2
%DemandTOTAL10 for demands rws 1 and 2
%CVaR for rows 1 and 2
% PI_TOTAL_EEV and LOAD_TOTAL_EEV and manually computed CVaR_Cost for final entry in Deter. row
\end{table}

\subsubsection{Results on IEEE RTS 24 Bus Test Case}
\label{subsubsection:Results on IEEE RTS 24 Bus Test Case}
\begin{figure}[ht]
\centering
\subfloat{\includegraphics[width=4.1cm]{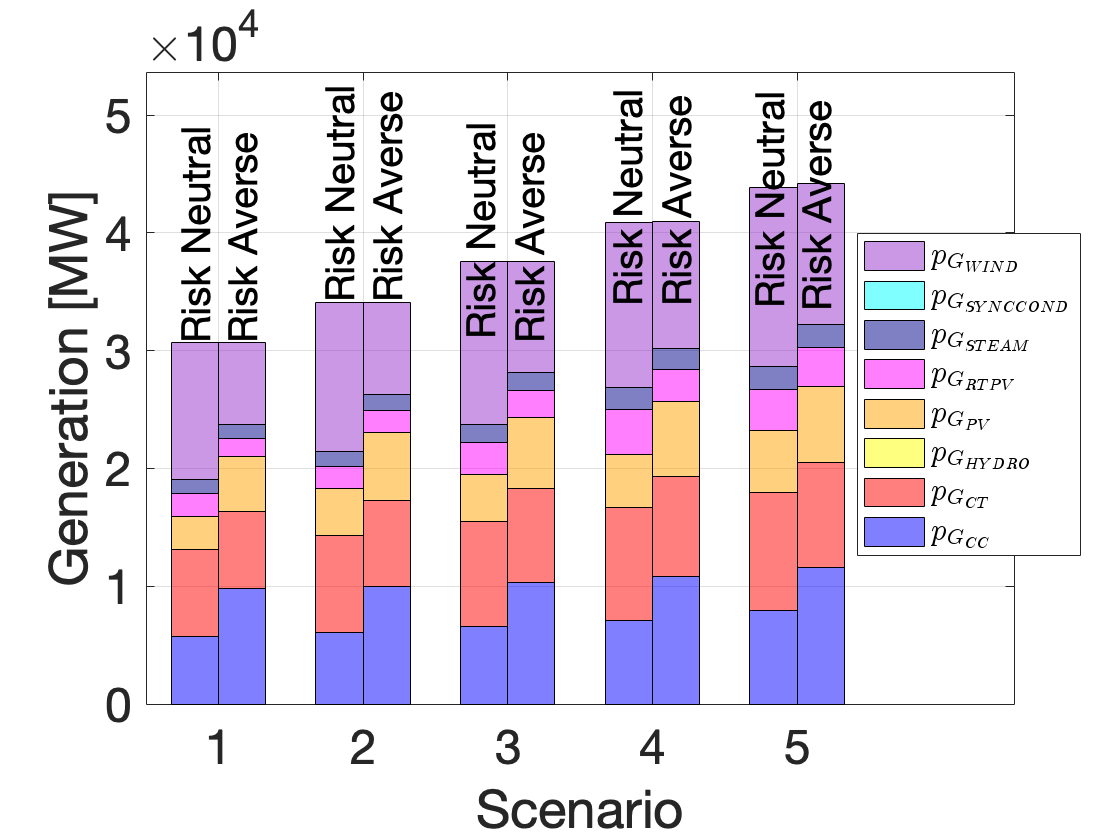}}
\hfil
\subfloat{\includegraphics[width=4.1cm]{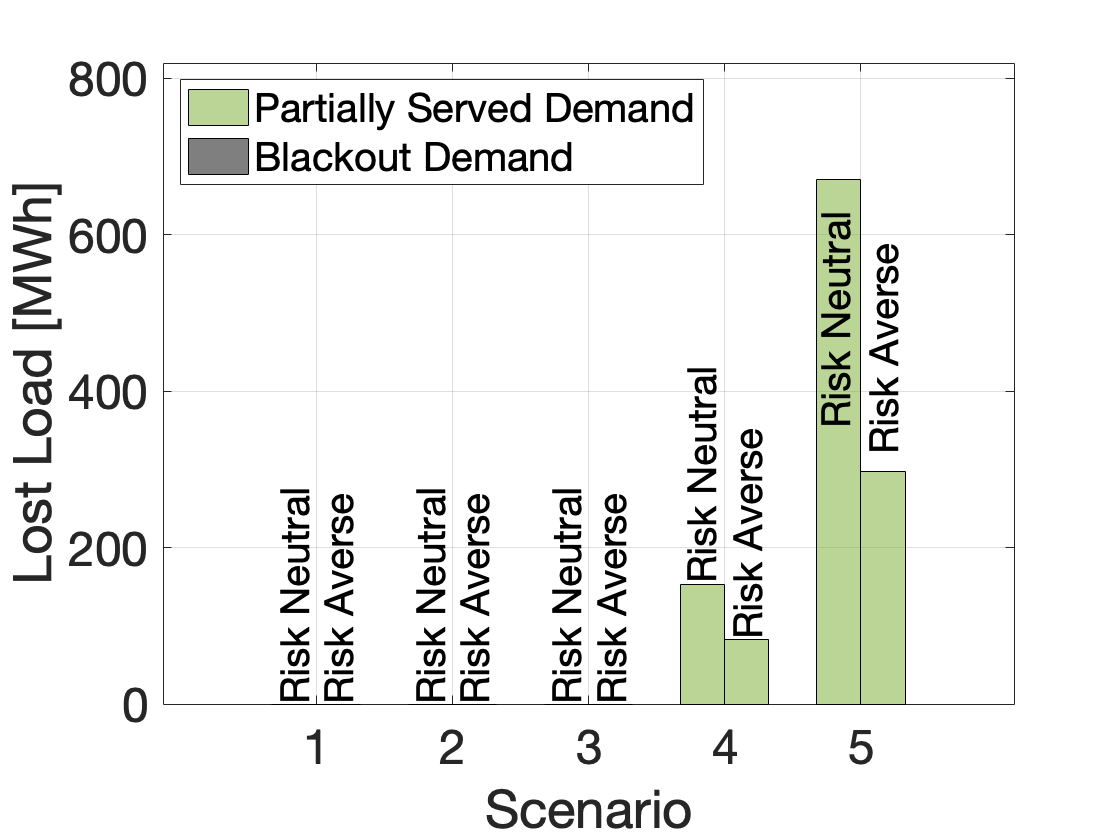}}
\hfil
\subfloat{\includegraphics[width=4.1cm]{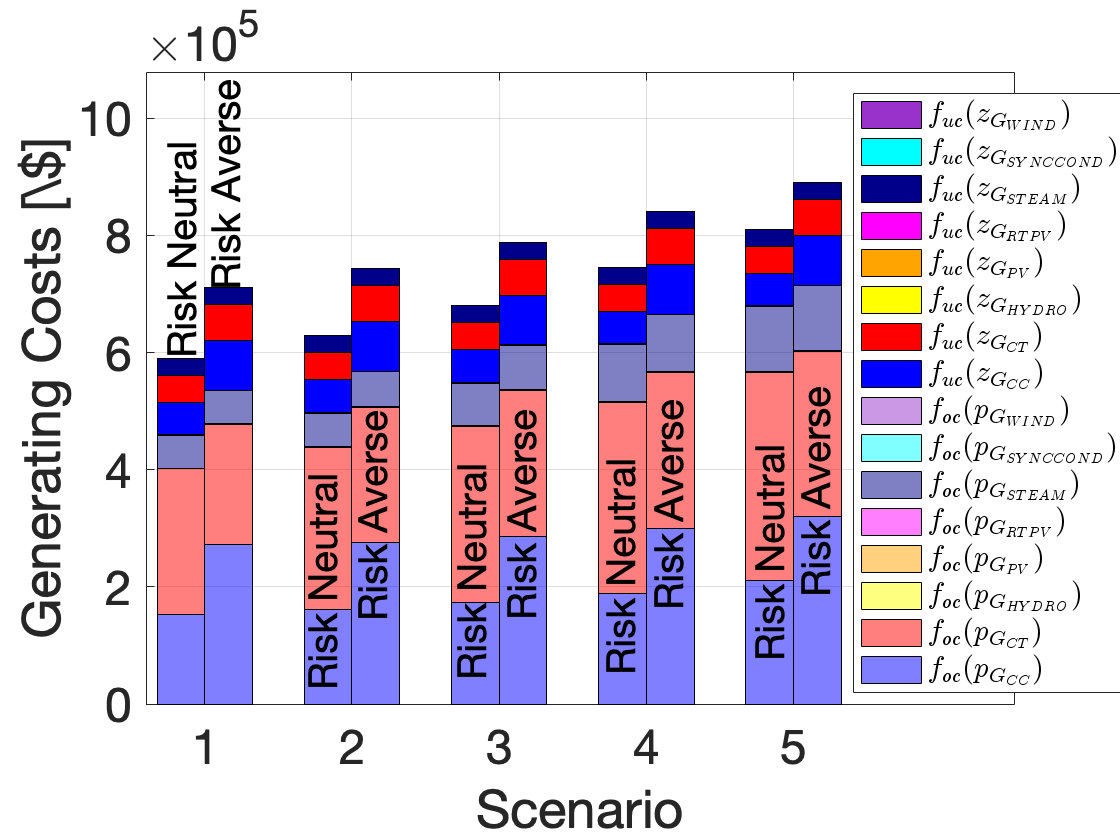}}
\hfil
\subfloat{\includegraphics[width=4.1cm]{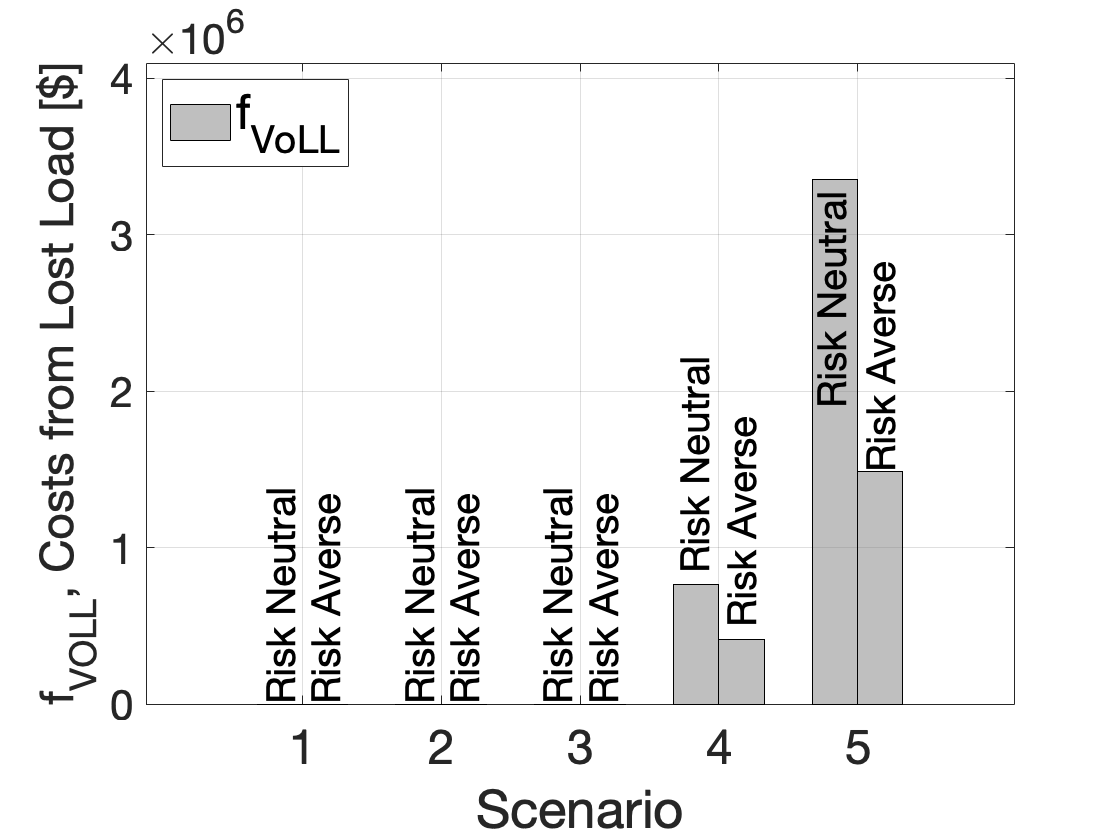}}
\caption{SPSPS (a) generation, (b) load shedding, (c) total costs (excluding VoLL), and (d) VoLL costs for the IEEE RTS 24-bus system optimized over 24 hours. Left bars are risk-neutral results for $\beta=0$ and right bars are risk-averse results for $\beta=0.995$. Subplots (a)-(d) together show that for the risk-averse strategy, higher production costs in scenarios 4 and 5 are overcome by savings in VoLL. The key difference in the commitment strategy between the risk-neutral and risk-averse strategies is that the risk-neutral strategy commits the (CC) generator at bus 23 instead of extra combustion turbine (CT) generators at bus 1. The combine-cycle generator at bus 23 provides more capacity and has a lower marginal cost but higher startup/shutdown costs than CT generators at buses 1 and 2. As a result, risk averse strategies lead to a larger maximum generation capacity and less lost load in the highest demand at every time step.} 
\label{fig:IEEE24BBarGenCostCompare14L}
%data: 
%script:
\vspace{-2em}
\end{figure}

In this section, we focus on the differences in costs between the risk-neutral and risk-averse cases for 14 active lines optimized over 24 hours. Figure \ref{fig:IEEE24BBarGenCostCompare14L}(a) shows that the risk-averse strategy spends more on operating and commitment costs in all scenarios. Despite spending more on operating costs,  Figure~\ref{fig:IEEE24BBarGenCostCompare14L}(a), (b), and (d) show that risk-averse decisions lead to more generation, less lost load, and costs associated with the lost load in the two highest demand scenarios.

\section{Conclusion}
\label{section:Conclusion}
% This paper presents a framework for system operators to carry out unit commitment during Public Safety Power Shut-offs. 
Both deterministic and stochastic frameworks are used to show how forecasts for high WFPI %vegetation flammability 
(that is often linked with occurrences of large wildfires) near transmission lines affect generator commitment and transmission line de-energization strategies. The proposed approach is implemented on the IEEE 14-bus test case and IEEE RTS 24 bus test cases. The optimal costs of commitment, operation, and lost load are compared to other public safety power shut-off models proposed in the literature. In general, our approach incurs less cost than previous methods that ignore economic costs, less system wildfire risks than planning methods unaware of wildfire risk, such as an $N$-$k$ approach, and it allows the operator to be more robust to changes in demand forecasts than deterministic approaches. Risk-averse decisions can be used by operators to reduce costs during high-demand periods of the wildfire season. However, additional studies may be needed to evaluate the fairness of blackouts that could result from risk-averse line de-energizations and the true costs of wildfire-related damages. The topology of the IEEE 24 bus system is a more realistic representation of the grid in the Southwest portion of the United States than the IEE 14 bus, yet it would be worth compiling and analyzing results on much larger transmission grids, such as the RTS GMLC or reduced 240-WECC feeder.


\begin{thebibliography}{00}
% SEction I: Introduction and Literature Review Citations
\bibitem{Muhs} J. W. Muhs, M. Parvania and M. Shahidehpour, "Wildfire Risk Mitigation: A Paradigm Shift in Power Systems Planning and Operation," in IEEE Open Access Journal of Power and Energy, vol. 7, pp. 366-375, 2020, doi: 10.1109/OAJPE.2020.3030023.

\bibitem{Arab} A. Arab, A. Khodaei, R. Eskandarpour, M. P. Thompson and Y. Wei, "Three Lines of Defense for Wildfire Risk Management in Electric Power Grids: A Review," in IEEE Access, vol. 9, pp. 61577-61593, 2021, doi: 10.1109/ACCESS.2021.3074477.

\bibitem{CPUC} “Public Safety Power Shutoffs,” Ca.gov, 2018. https://www.cpuc.ca.gov/psps. (accessed Jan. 20, 2025).

% California Public Utilities Commission. Public Safety Power Shut-
% off (PSPS)/De-Energization. Accessed: Oct. 2020. [Online]. Available:
% https://www.cpuc.ca.gov/deenergization

\bibitem{Abatzoglou} J. T. Abatzoglou, C. M. Smith, D. L. Swain, T. Ptak, and C. A. Kolden, "Population exposure to pre-emptive de-energization aimed at averting wildfires in Northern California," Environ. Res. Lett., vol. 15, no. 9, Aug. 2020, Art. no. 094046.

\bibitem{EPA} “Climate Change Indicators: Wildfires | US EPA,” US EPA, Jul. 2016. https://www.epa.gov/climate-indicators/climate-change-indicators-wildfires (accessed Jan. 20, 2025).

% “Climate Change Indicators: Wildfires.” EPA, Environmental Protection Agency, https://www.epa.gov/climate-indicators/climate-change-indicators-wildfires.

\bibitem{UN} K. Rukikaire,“Number of wildfires to rise by 50 per cent by 2100 and governments are not prepared, experts warn,” UN Environment, 2022. https://www.unep.org/news-and-stories/press-release/number-wildfires-rise-50-cent-2100-and-governments-are-not-prepared (accessed Jan. 20, 2025).

% “Number of Wildfires to Rise by 50 \% by 2100 and Governments Are Not Prepared, Experts Warn.” UN Environment, 23 Feb. 2022, https://www.unep.org/news-and-stories/press-release/number-wildfires-rise-50-2100-and-governments-are-not-prepared.

\bibitem{Rhodes} N. Rhodes, L. Ntaimo and L. Roald, "Balancing Wildfire Risk and Power Outages Through Optimized Power Shut-Offs," in IEEE Transactions on Power Systems, vol. 36, no. 4, pp. 3118-3128, July 2021, doi: 10.1109/TPWRS.2020.3046796.

\bibitem{Kody} A. Kody, A. West and D. K. Molzahn, "Sharing the Load: Considering Fairness in De-energization Scheduling to Mitigate Wildfire Ignition Risk using Rolling Optimization," 2022 IEEE 61st Conference on Decision and Control (CDC), Cancun, Mexico, 2022, pp. 5705-5712, doi: 10.1109/CDC51059.2022.9993295.


% A. Kody, R. Piansky, and D. Molzahn, “Optimizing Transmission
% Infrastructure Investments to Support Line De-energization for Mitigating Wildfire Ignition Risk,” in 11th Bulk Power Systems Dynamics
% and Control Symposium (IREP XI), Banff, Canada, July 2022.

\bibitem{Rhodes2} N. Rhodes, C. Coffrin, and Line Roald, “Security constrained optimal power shutoff for wildfire risk mitigation,” IET Generation Transmission \& Distribution, vol. 18, no. 18, pp. 2972–2986, Aug. 2024, doi: https://doi.org/10.1049/gtd2.13246.

% N. Rhodes and C. Coffrin and L. Roald, "Security Constrained Optimal Power Shutoff", arXiv Preprint 2304.13778, 2023.

\bibitem{Kody2} A. Kody, A. West, and D. Molzahn, “Sharing the Load: Considering Fairness in De-energization Scheduling to Mitigate Wildfire Ignition Risk using Rolling Optimization.” Accessed: Oct. 24, 2022. [Online]. Available: https://arxiv.org/pdf/2204.06543.pdf‌

% \bibitem{Astudillo} A. Astudillo, B. Cui and A. S. Zamzam, "Managing Power Systems-Induced Wildfire Risks Using Optimal Scheduled Shutoffs," 2022 IEEE Power \& Energy Society General Meeting (PESGM), Denver, CO, USA, 2022, pp. 1-5, doi: 10.1109/PESGM48719.2022.9917206.

\bibitem{Umunnakwe} A. Umunnakwe, M. Parvania, H. Nguyen, and J. D. Horel, and K. R. Davis, "Data-driven spatio-temporal analysis of wildfire risk to power systems operation," 2021.

\bibitem{Haimes} Y. Haimes, L. Lasdon  and D.  Wismer, On a bicriterion formulation  of  the  problems  of  integrated  system identification  and  system  optimization.  IEEE Transactions  on  Systems,  Man,  and  Cybernetics, 1:296-297, 1971.

\bibitem{Bayani} R. Bayani, M. Waseem, S. D. Manshadi and H. Davani, "Quantifying the Risk of Wildfire Ignition by Power Lines Under Extreme Weather Conditions," in IEEE Systems Journal, vol. 17, no. 1, pp. 1024-1034, March 2023, doi: 10.1109/JSYST.2022.3188300.

\bibitem{Moreno} R. Moreno et al., "Microgrids Against Wildfires: Distributed Energy Resources Enhance System Resilience," in IEEE Power and Energy Magazine, vol. 20, no. 1, pp. 78-89, Jan.-Feb. 2022, doi: 10.1109/MPE.2021.3122772.

%\bibitem{Trakas1} D. N. Trakas and N. D. Hatziargyriou, "Resilience Constrained Day-Ahead Unit Commitment Under Extreme Weather Events," in IEEE Transactions on Power Systems, vol. 35, no. 2, pp. 1242-1253, March 2020, doi: 10.1109/TPWRS.2019.2945107.

\bibitem{Trakas2} D. N. Trakas and N. D. Hatziargyriou, "Optimal Distribution System Operation for Enhancing Resilience Against Wildfires," in IEEE Transactions on Power Systems, vol. 33, no. 2, pp. 2260-2271, March 2018, doi: 10.1109/TPWRS.2017.2733224.

%\bibitem{Fobes} D. M. Fobes, H. Nagarajan and R. Bent, "Optimal Microgrid Networking for Maximal Load Delivery in Phase Unbalanced Distribution Grids: A Declarative Modeling Approach," in IEEE Transactions on Smart Grid, 2022, doi: 10.1109/TSG.2022.3208508.

% Section II: Public Safety Power Shut-offs During Day Ahead and Real-Time Operation
%\bibitem{SDGEPSPS} C. Arends, A. Llacuna, M. Freels, et. al., "PSPS Model Updatesm," 2022 California Public Utilities Commission Workshop on Public Safety Power Shut-offs, April 18, 2023.

%Section III: Modeling PSPS for WF Risk Mitigations
%III:E Line constraints
% \bibitem{WFPI} Wildland Fire Potential Index (WFPI) | U.S. Geological Survey. (n.d.). https://www.usgs.gov/fire-danger-forecast/wildland-fire-potential-index-wfpi

\bibitem{WFPI} B. Sickmeyer, “Fire Danger Map and Data Products,” Usgs.gov, 2025. https://firedanger.cr.usgs.gov/apps/staticmaps (accessed Jan. 20, 2025).

% Wildland Fire Potential Index (WFPI) Fire Danger Maps and Products Page, U.S. Geological Survey. (n.d.). https://firedanger.cr.usgs.gov/apps/staticmaps

\bibitem{Bienstock} D. Bienstock, A. Verma, "The N-k Problem in Power Grids: New Models, Formulations, and Numerical Experiments", SIAM Journal on Optimization, vol. 20, no. 5, pp. 2352-2380, 2010, doi = 10.1137/08073562X.

% III.G.1 Two Stage Stochastic PSPS

\bibitem{Growe-Kuska} N. Gr\"owe-Kuska, H. Heitsch, and W. R\"omisch, “Scenario reduction and scenario tree construction for power management problems,” in 2003 IEEE Bologna PowerTech - Conference Proceedings, 2003, vol. 3, pp. 152–158.

\bibitem{Noyan} N. Noyan, "Risk-averse two-stage stochastic programming with an application to disaster management," Computers \& Operations Research, vol. 39, no. 3, pp. 541-559, 2012. doi:10.1016/j.cor.2011.03.017

\bibitem{Rockafellar} R. T. Rockafellar, S. Uryasev, "Optimization of Conditional Value at Risk," Journal of Risk, 2000, vol. 2, pp. 21-41.

\bibitem{Shaked} M. Shaked, J.G. Shanthikumar, "Stochastic orders and their applications," Boston: AssociatedPress, 1994.

%\bibitem{Royset} T. R. Rockafellar, J. O. Royset, "Engineering Decisions Under Risk Averseness," Naval Postgraduate School Office Department of Operations Research, Monterrey, CA, 2014, https://api.semanticscholar.org/CorpusID:201033288

\bibitem{Rawlings} 
J. B. Rawlings, D. Q. Mayne, M. Diehl, "Model Predictive Control: The Theory, Computation, and Design, 2nd ed. Madison, Wisconsin, Nob Hill Publishing, 2017.

% J. B. Rawlings, D. Q. Mayne, M. M. Diehl, "Model Predictive Control: The Theory, Computation, and Design 2nd Edition," Nob Hill Publishing, 2017, ISBN: 0975937731

% IV.A. Results: Data and Test Cases 

\bibitem{RTSGMLC} C. Barrows, A. Bloom, A. Ehlen, J. Ik¨aheimo, J. Jorgenson, D. Krishnamurthy, J. Lau, B. McBennett, M. O’Connell, E. Preston et al., “The IEEE Reliability Test System: A Proposed 2019 Update,” IEEE Transactions on Power Systems, vol. 35, no. 1, pp. 119–127, 2019.
% "Reliability Test System Grid Modernization Lab Consortium (RTS-GMLC) Transmission System Database," Grid Modernization Lab, (2022). https://github.com/GridMod/RTS-GMLC

%\bibitem{IEEE14} “IEEE 14-Bus System - Illinois Center for a Smarter Electric Grid (ICSEG),” Illinois.edu, 2013. https://icseg.iti.illinois.edu/ieee-14-bus-system/ (accessed Feb. 15, 2021).

%\bibitem{PSCAD} "IEEE 14 bus technical note" Manitoba HVDC Research Centre (PSCAD), December 30th, 2014.

% \bibitem{WFPI2} Wildland Fire Potential Index (WFPI) Fire Danger Maps and Products Page, U.S. Geological Survey. (n.d.). https://firedanger.cr.usgs.gov/apps/staticmaps

\bibitem{PowerGridLibIEEE14} S. Babaeinejadsarookolaee, A. Birchfield, R. D. Christie, C. Coffrin, C. L. DeMarco, R. Diao, M. Ferris, S. Fliscounakis, S. Greene, R. Huang, C. Josz, R. Korab, B. C. Lesieutre, J. Maeght, D. K. Molzahn, T. J. Overbye, P. Panciatici, B. Park, J. Snodgrass, and R. D. Zimmerman, “The Power Grid Library for Benchmarking AC Optimal Power Flow Algorithms,” August 2019, arXiv:1908.02788.

% "OPF Case 14 Bus," Benchmark IEEE PES Power Grid Library for Optimal Power Flow v23.07, July 23, 2023: https://github.com/power-grid-lib/pglib-opf/blob/master/pglib\_opf\_case14\_ieee.m

\bibitem{Bresenham} J. E. Bresenham, “Algorithm for computer control of a digital plotter,” ACM eBooks, pp. 1–6, Jul. 1998, doi: https://doi.org/10.1145/280811.280913.

% "ICS183: Bresenham's algorithm," CS 112: Computer Graphics, University California Irvine Information and Computer Science Department, https://ics.uci.edu/~gopi/CS112/web/handouts/OldFiles/Bresenham.pdf

\bibitem{Mohagheghi} S. Mohagheghi and S. Rebennack, “Optimal resilient power grid operation during the course of a progressing wildfire,” Int. J. Elect. Power Energy Syst., vol. 73, pp. 843–852, 2015.

\bibitem{Farzin} H. Farzin, M. Fotuhi-Firuzabad, and M. Moeini-Aghtaie, “Stochastic energy management of microgrids during unscheduled islanding period,” IEEE Trans. Ind. Informat., vol. 13, no. 3, pp. 1079–1087, Jun. 2017.

% IV.C. Results: Stochastic PSPS 

\bibitem{Sarykalin} S. Sarykalin, G. Serraino, S. Uraysev, "value-at-Risk vs. Conditional Value-at-Risk Management and Optimization," Tutorials in Operations Research, 2008: 10.1287/educ.1080.0052

% \bibitem{Wasserstein} Wikipedia Contributors, “Wasserstein metric,” Wikipedia, Jul. 09, 2024. https://en.wikipedia.org/wiki/Wasserstein\_metric\#cite\_note-3 (accessed Sep. 25, 2024).

% \bibitem{Heitsch} H. Heitsch, W. R¨ omisch, A note on scenario reduction for two-stage stochastic programs, Operations Research Letters, 35(6):731-738, 2007

% \bibitem{Ryan} S. M. Ryan, R. J. . -B. Wets, D. L. Woodruff, C. Silva-Monroy and J. -P. Watson, "Toward scalable, parallel progressive hedging for stochastic unit commitment," 2013 IEEE Power \& Energy Society General Meeting, Vancouver, BC, Canada, 2013, pp. 1-5, doi: 10.1109/PESMG.2013.6673013.

% \bibitem{Ordoudis} C. Ordoudis, P. Pinson, M. Zugno and J. M. Morales, "Stochastic unit commitment via Progressive Hedging — extensive analysis of solution methods," 2015 IEEE Eindhoven PowerTech, Eindhoven, Netherlands, 2015, pp. 1-6, doi: 10.1109/PTC.2015.7232629.

% \bibitem{Deb} K. Deb, Multi-Objective Optimization using Evolutionary Algorithms, John Wiley \& Sons, Inc., 2001.


\end{thebibliography}
\end{document}